\documentclass[prd,aps,10pt,nofootinbib]{revtex4-2}
\usepackage{graphicx,amsmath,amssymb} 
\usepackage{xcolor}
\usepackage{verbatim}
\usepackage{physics,float}
\usepackage{hyperref}
\usepackage{cancel}
\hypersetup{
	colorlinks=true,
	linkcolor=red,
	citecolor=blue,			
	urlcolor=blue,			
	filecolor=blue}
\usepackage{makecell} 
\begin{document}


\title{Index theorem with Minimally Doubled Fermions\\in four space-time dimensions}
\author{Abhijeet Kishore$^1$}
\email{akishore@iitk.ac.in}

\author{Subhasish Basak$^2$}
\email{sbasak@niser.ac.in}

\author{Dipankar Chakrabarti$^1$}
\email{dipankar@iitk.ac.in}

\affiliation{
$^1$Department of Physics, Indian Institute of Technology Kanpur,
Kanpur-208016, India}

\affiliation{$^2$School of Physical Sciences, National Institute of
Science Education and Research, An OCC of Homi Bhabha National
Institute, Jatni-752050, India}

\date{\today}
\begin{abstract}
We determine the zero eigenmode spectrum of
Minimally Doubled
Fermions (MDF), namely in Karsten-Wilczek (KW) and Borici-Creutz (BC)
formulations on the 4-dimensional space-time lattice. We employ background gauge fields with integer valued topological charges. The Atiyah-Singer index theorem is verified in the presence of two different background gauge fields, namely Smit-Vink \cite{Smit:1986fn} and cooled down  MILC asqtad ensembles with $N_f=2+1$ dynamical
flavors of quarks~\cite{Bernard:2001av}.
Using flavored mass terms~\cite{Creutz:2010bm,Durr:2022mnz}, we find
that the spectral flow of the eigenvalues detects the topology
of the background gauge field.  With the use of the modified
chirality operator, we obtain chiralities of the zero eigenmodes and the fermionic
topological charge.
\end{abstract}

\maketitle

\section{Introduction}\label{sec:intro}
Eigenvalue spectra of Dirac operators reflect QCD dynamics such as
Bank-Casher relation that relates the zero eigenvalue density,
\textit{i.e.}, zero eigenmodes, to the chiral condensate. To obtain
and identify zero-eigenmodes, the chiral symmetry needs to be obeyed
in the lattice fermion formulation. The two well-known chiral formulations
of lattice fermions are overlap \cite{Narayanan:1997fx,Edwards:1998yw,
Edwards:1998vu,Edwards:1999yi,Adams:1998eg} and domain wall
\cite{Kaplan:1992bt,Shamir:1993zy,Blum:2001qg,Blum:2000kn,Fukaya:2010na,
Brower:2012vk,Cossu:2016eqs}.
The study of  eigenspectra, topology and chiral condensates of
these two, along with staggered and Wilson fermions, have been
extensively carried out and used to explore the non-perturbative
dynamics of QCD.

The continued search for alternative lattice chiral fermions that would
be numerically cheaper for QCD simulations led to the proposal of
minimally doubled fermions (MDF) with local action
that preserves chiral symmetry on the lattice. Two relatively popular
implementations of MDF that are mostly explored are those by
Karsten-Wilczek (KW) \cite{Karsten:1981gd,Wilczek:1987kw} and
Borici-Creutz (BC) \cite{Creutz:2007af,Borici:2007kz}. A string of
investigations on MDF, particularly KW and BC fermions, have been
carried out focusing on mixed action spectroscopy \cite{Basak:2017oup,
Godzieba:2024uki,Weber:2013tfa,Weber:2015hib,Weber:2016dgo},
eigenspectra \cite{Durr:2020yqa}, taste structures \cite{Borsanyi:2025big,
Ammer:2024yro,Durr:2024ttb,WeberCombined}, chiral symmetry breaking
\cite{Osmanaj:2018pqb}, and index theorem \cite{Creutz:2010bm,
Durr:2022mnz}. Studies on formal aspects include renormalization
properties \cite{Capitani:2010nn,Vig:2024umj,Capitani:2013zta,
Capitani:2013fda,Capitani:2009ty,Capitani:2009yn,Capitani:2010ht,
Osmanaj:2022mzs,Kimura:2011ik}, phase structure \cite{Misumi:2012uu,
Kimura:2012df,Misumi:2012ky}, anomaly \cite{Tiburzi:2010bm}, and
construction of chiral perturbation theory \cite{Shukre:2025bnm,
WeberCombined}. Besides, there is a recent attempt to perform dynamical
simulation with KW fermions \cite{Vig:2024umj}.

In this paper, we attempt to investigate the 
Atiyah-Singer index theorem \cite{Atiyah:1963zz} and
the universality of near-zero eigenmodes in sectors with  different topological
charges \cite{Shuryak:1992pi,Leutwyler:1992yt}. In the presence of 
background gauge field with an integer valued topological charge, the
chiral fermions are expected to satisfy the Atiyah-Singer index theorem.
KW and BC fermions are shown to satisfy the index theorem in two
dimensions \cite{Chakrabarti:2009sa, Creutz:2010bm,Durr:2022mnz,
Pernici:1994yj,Tiburzi:2010bm}. Here we extend that study to four
dimensions. We have constructed four dimensional $SU(3)$ background gauge fields
with a fixed topological charge following the prescription of Smit and Vink
\cite{Smit:1986fn,Gattringer:1997c} and observed the flow of eigenvalues
as the mass varies. The index of the Hermitian Dirac
operator thus obtained from the spectral flow gives us the difference
between number of positive and negative chirality zero eigenmodes. For
the measurement of the chirality of the zero eigenmodes, we have worked
with modified chirality operator and used flavored mass terms
\cite{Creutz:2010bm,Durr:2022mnz}. To verify the universality of the
zero eigenmodes, we repeated the study with dynamical QCD configurations
using the publicly available MILC $N_f=2+1$ asqtad lattices
that are subjected to cooling \cite{Bonnet:2001rc} in order to
accurately identify various topological charges. This paper is built
on and extends our previous work presented in \cite{Kishore:2025fxt}.

This paper is organized as follows: following this Introduction, in
Sec. \ref{sec:fermion} we write down both the KW and BC action in
momentum space and point out the doubler species. In Sec.
\ref{sec:specflow}, we introduce the spectral flow and define the
flavored mass. We also describe the generation of background gauge
fields with specific topological charge. Subsequently, we study the
flow of the eigenvalues against the variation of
mass thereby determining the index of the Dirac operator.
Following, in Sec. \ref{sec:index_under_general_gauge_field}, using
MILC's dynamical guage field configurations we employ cooling to
identify various topological charge configurations and repeat our
spectral flow analysis to reproduce the Index theorem. Finally, in Sec.
\ref{sec:conclusion}, we conclude with a discussion of our results,
along with a summary. We present some results for a different topological charge
$Q$ with MILC lattices in Appendix \ref{appendix_A}, and some calculations
pertaining to the eigenvalues of the MDF Dirac operator in Appendix \ref{appendix_B}.

\section{Karsten-Wilczek and Borici-Creutz fermions}\label{sec:fermion}

According to the Nielsen-Ninomiya theorem \cite{Nielsen:1981hk}, fermion
actions having exact chiral symmetry must have even number of doublers
on the lattice. The doublers are extra fermionic species arising because
of discretized space-time. The minimum number of doublers on the lattice
is two and such fermions are generally referred to as Minimally
Doubled Fermions (MDF). The Karsten-Wilczek (KW) \cite{Karsten:1981gd,
Wilczek:1987kw} and Borici-Creutz (BC) \cite{Creutz:2007af,Borici:2007kz}
fermions are the two popular members of MDF family. In addition to the
naive discretization of the continuum fermion action, both KW and BC
contain terms that anticommute with $\gamma_5$, thereby keeping the actions
chiral. In Wilson action, a mass-like discretized 4-dimensional
Laplacian is added to the naive fermion action that breaks all the
chiral symmetries. In contrast, KW fermion multiplies the spatial part
of Laplacian by a factor of $i\gamma_4$. It breaks the hypercubic
symmetry, but preserves rotational symmetry along the temporal axis,
thereby retains chiral symmetry of the lattice action. In 4-dimension,
the KW action \cite{Capitani:2010ht} in the presence of a gauge field
$U_{\mu}(x)$ can be written as
\begin{equation}
\begin{split}
S_{\text{KW}} = \sum_{\substack{x}}\Big[ \frac{1}{2} \sum_{\substack{
\mu=1}}^{4} \bar{\psi}(x) \gamma_{\mu} \big\{ U_{\mu}(x) \psi(x+\hat{\mu})
- U^{\dagger}_{\mu}(x-\hat{\mu}) \psi(x-\hat{\mu}) \big\} + m\bar{\psi}
(x) \psi(x) \\
- \frac{i}{2} \sum_{\substack{j=1}}^{3} \bar{\psi}(x) \gamma_{4}
\big\{ U_{j}(x) \psi(x+\hat{j}) - 2\psi(x)  + U^{\dagger}_{j}(
x-\hat{j}) \psi(x-\hat{j})\big\}  \Big].
\end{split}    \label{kw_action}
\end{equation}
The two doubler species become manifest in the Brillouin zone when
expressed in momentum space. For simplicity, we consider the
free massless KW Dirac operator,
\begin{eqnarray}
D_{\text{KW}}(p) &=& \sum_{\substack{\mu=1}}^{4} i\gamma_{\mu} \,
\sin p_{\mu} + 2i \gamma_4 \sum_{\substack{j=1}}^{3} \big[\sin (p_{j}/2)
\big]^2. \label{kw_mom_dirac}
\end{eqnarray} 
The poles of the propagator $D_\text{KW}^{-1}(p)$ are found to be at $p=(0,0,
0,0)$ and $(0,0,0,\pi)$ corresponding to the two doublers or tastes,
as it is often called.

In BC fermion, the discretized Laplacian is multiplied by a factor of
$-i\gamma^\prime_{\mu}$ which too breaks the hypercubic symmetry, but
preserves the rotational symmetry along the axis joining the origin
and $(\frac{\pi}{2},\frac{\pi}{2},\frac{\pi}{2},\frac{\pi}{2})$,
retaining the chiral symmetry of the lattice action. In 4-dimension,
the BC action \cite{Capitani:2010ht} in the presence of a gauge field
$U_{\mu}(x)$ can be written as
\begin{equation}
\begin{split}
S_{\text{BC}} = \sum_{\substack{x}}\Big[ \frac{1}{2} \sum_{\substack{
\mu=1}}^{4} \bar{\psi}(x) \gamma_{\mu} \big\{ U_{\mu}(x) \psi(x+\hat{\mu})
- U^{\dagger}_{\mu}(x-\hat{\mu}) \psi(x-\hat{\mu}) \big\}  + m\bar{\psi}
(x) \psi(x) \\
+ \frac{i}{2} \sum_{\substack{\mu=1}}^{4} \bar{\psi}(x) \gamma'_{\mu}
\big\{ U_{\mu}(x) \psi(x+\hat{\mu}) - 2\psi(x)  + U^{\dagger}_{\mu}
(x-\hat{\mu}) \psi(x-\hat{\mu})\big\} \Big],
\end{split} \label{bc_action}
\end{equation}
where $\gamma^\prime_{\mu}=\Gamma - \gamma_{\mu}$, $\;\Gamma=(1/2) \sum_\mu
\gamma_{\mu}$. Similar to the KW case, we can expose the doublers by
working with a free massless theory in momentum space. The corresponding BC Dirac
operator is
\begin{eqnarray}
D_{\text{BC}}(p) &=& \sum_{\substack{\mu}} i
\gamma_{\mu} \, \sin p_{\mu} - 2i \sum_{\substack{\mu}} \gamma'_{\mu} \,
\big[\sin (p_{\mu}/2) \big]^2.  \label{bc_mom_dirac}
\end{eqnarray}
From $D_\text{BC}^{-1}(p)$, it is evident that the poles are at
$p=(0,0,0,0)$ and $(\frac{\pi}{2},\frac{\pi}{2},\frac{\pi}{2},\frac{\pi}
{2})$. From Eqs. (\ref{kw_mom_dirac}, \ref{bc_mom_dirac}) it follows
that KW and BC Dirac operators both satisfy $\gamma_5$-hermiticity,
$D^{\dagger}=\gamma_5 D \gamma_5$ and are diagonal in the momentum space.

\section{Spectral flow and index}\label{sec:specflow}

\begin{figure}[htb]
\makebox[\textwidth]{\includegraphics[width=0.45\textwidth]{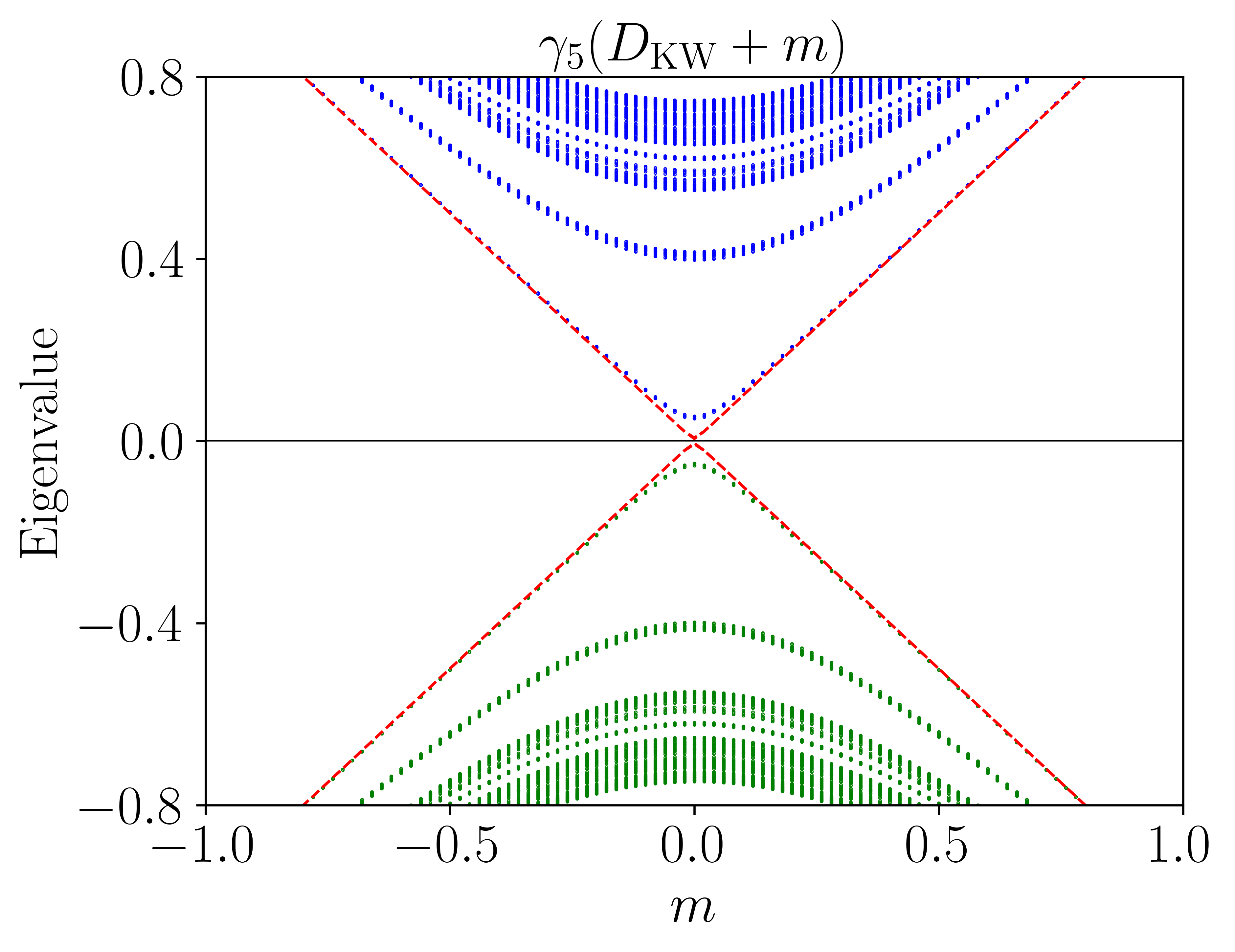} \hspace{0.1in}
\includegraphics[width=0.45\textwidth]{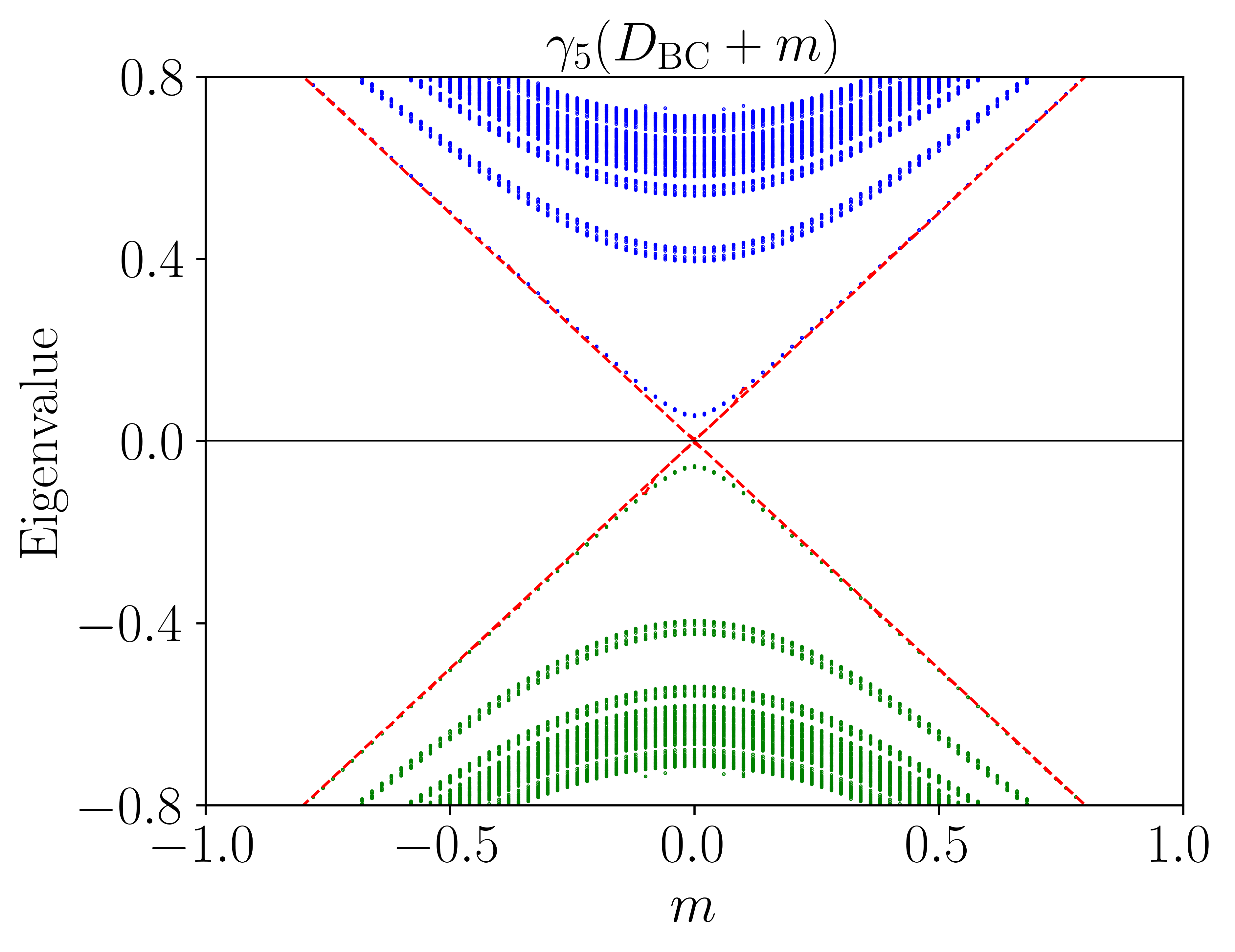}}
\vspace{-0.2in}
\caption{Spectral flow with respect to degenerate bare masses $m$ for $\pm$ chiral states for 
both KW and BC fermions under $Q=-2$ and $\delta=0.05$ background $8^3\times8$
Smit-Vink lattice. Eigenvalues in left and right panels correspond to $\gamma_5(D_{\text{KW}}+m)$ and
$\gamma_5(D_{\text{BC}}+m)$, respectively. The color coding is intended to guide the eye.}
\label{fig:normal_mass_specflow}
\end{figure}

In the continuum, the Atiyah-Singer index theorem relates the difference in
the number of negative and positive chirality zero modes of massless
Dirac operator to the integer valued topological charge $Q$
in four space-time dimensions
\cite{Atiyah:1963zz}
\begin{equation}
\text{index}(D) = n_{+} - n_{-} = Q. \label{qtop}
\end{equation}
The zero modes are eigenstates
of the Dirac operator with zero eigenvalues. A large body of studies
with Wilson and staggered fermions \cite{Smit:1986fn,Smit:1987zh,
Itoh:1987iy,Neuberger:1997fp,Adams:2009eb,Follana:2004sz,Follana:2005km,
Azcoiti:2014pfa,SWME:2020yip} have shown the index theorem to hold on
the lattice too. It is also studied for MDF in 2-dimension \cite{Creutz:2010bm,
Durr:2022mnz}. An approach to the index theorem on the lattice, away from
the continuum, is to consider a hermitian version of the Dirac operator, $H(m)
= \gamma_5 \,(D + m)$, and observe the spectral flow of the eigenvalues
of $H(m)$ in a background gauge field. The flow relates the eigenvalues
with the topological charge. As the mass varies, the
eigenvalue flow shows that those corresponding to the zero-modes cross the
origin with slopes $\pm$, depending on their $\pm$ chiralities, while the
non-zero modes do not exhibit any zero-crossing (please refer to Appendix \ref{appendix_B}
particularly Eq. (\ref{eq:exact_h_eigenvalue})). The index of the Dirac
operator is derived from the total number of eigenvalue zero-crossings.
The eigenmodes of $H$ can be derived from that of $D$. If $\ket{\psi}$
is an eigenvector of $D$ with eigenvalue $i\lambda$, then $\gamma_5
\ket{\psi}$ is also an eigenvector of $D$ with eigenvalue $-i\lambda:$
\begin{equation}
D \ket{\psi} = i \lambda \ket{\psi} \implies D(\gamma_5 \ket{\psi})
= -i \lambda (\gamma_5 \ket{\psi}), \hspace{0.2in} \lambda \in
\mathbb{R}. \label{evDg5D}
\end{equation}
Hence, a zero eigenmode $\ket{\psi}$ of $D$ with $\pm$ chirality is also
an eigenmode of $H$ with eigenvalue $\pm m$, and therefore crosses the origin
with $\pm$ slope as $m$ is varied. For Dirac operators with doubler species having
degenerate masses, there is an equal number of positive chirality $(n_+)$
and negative chirality $(n_-)$ states. Thus, the index($D$) in such cases
turns out to be zero and the spectral flow obtained for KW and BC
fermions shown in the Fig.
\ref{fig:normal_mass_specflow} bears out this fact. The gauge field,
used for the flow shown, is generated using Smit-Vink prescription
discussed in the following subsection.

Connecting the index($D$) with the topological charge seemingly requires
careful consideration of the doublers' masses and a better measure of
chirality of the zero-modes. We will take up this issue later in the
section \ref{subsec:index_flav_mass}.

\subsection{Gauge configuration with fixed \texorpdfstring{$Q$}{Q\_top}}\label{sec:gauge_config_with_tailor_made_qtop}

For our study of the index theorem on the lattice with MDF, we generate background
gauge configuration with a fixed topological charge $Q$. Here,
we follow the prescription laid out by Smit and Vink in \cite{Smit:1986fn,
Gattringer:1997c} that results in a constant field strength tensor
$F_{\mu \nu}(x)$. In terms of $F_{\mu\nu}$, the topological
charge on the lattice is defined as
\begin{eqnarray}
Q = \sum_x q(x) &=& -\frac{1}{32\pi^2} \sum_x
\epsilon_{\mu\nu\rho\sigma}\, \text{Tr}\left(F_{\mu \nu} F_{\rho \sigma}
\right) \nonumber \\
 &=& -\frac{1}{4\pi^2} \sum_x \text{Tr}[F_{12}F_{34} - F_{13}F_{24}
+ F_{23}F_{14}]. \label{topo_charge}
\end{eqnarray}
The lattice sites are $x_\mu \in [0,L_{\mu}]$ and here we have
assumed $L_\mu=L \; \forall \;\mu$ where $L$ is the number of sites
in each direction, thus $x_\mu = 0,1, \ldots, (L-1)$. The corresponding
link variables $U_{\mu}(x) \in\;$ SU(3) are
\begin{eqnarray}
U_1(x) &=& \exp(-i\omega_1 x_2 \tau_k) \;\;\text{and} \;\; U_2(x) =
\begin{cases}
1 & \text{for} \;\; x_2 = 0,1, \ldots, (L-2) \\ 
\exp(i\omega_1 L x_1 \tau_k) & \text{for} \;\; x_2=L-1
\end{cases}, \label{u1u2} \\
U_3(x) &=& \exp(-i\omega_2 x_4 \tau_k) \;\; \text{and} \;\; U_4(x) =
\begin{cases}
1 & \text{for} \;\; x_4 = 0,1, \ldots,(L-2) \\
\exp(i\omega_2 L x_3 \tau_k) & \text{for} \;\; x_4=L-1
\end{cases}, \label{u3u4}
\end{eqnarray}
where $\tau_k$ is any one of the Gell-Mann matrices with $k=1,2,
\ldots, 7$. The $\omega_{1,2}$ are constant field strength
given by
\begin{equation}
\omega_{1,2} = \frac{2\pi\,n_{1,2}}{L^2} \;\;\;\text{where} \;\;
n_{1,2} \in \mathbb{Z}. \label{omg}
\end{equation}
The gauge links thus defined, the elementary plaquettes become constant,
\begin{eqnarray}
U_{\mu \nu}(x) &=& U_{\mu}(x)U_{\nu}(x+\hat{\mu})U^{\dagger}_{\mu}(x+
\hat{\nu})U^{\dagger}_{\nu}(x), \nonumber \\
U_{12}(x) &=& \exp(i\omega_1\tau_k), \;\;\;U_{34}=\exp(i\omega_2 \tau_k)
\;\;\;\text{and the rest} \;\; U_{\mu \nu}(x)=1. \label{uPconst}
\end{eqnarray}
The field tensor $F_{\mu \nu}$ extracted from the above elementary
plaquette $U_{\mu\nu}$ in (\ref{uPconst}) yields,
\begin{eqnarray}
F_{\mu \nu}(x) &=& \frac{1}{2i} \big[U_{\mu \nu}(x) - U_{\mu \nu}^\dagger
(x) \big]_\text{AH} \label{fmunuAH} \\
\Rightarrow \;\; F_{12} &=& \tau_k\,\sin(\omega_1), \;\;\;F_{34} = \tau_k
\,\sin(\omega_2) \;\;\;\text{and the rest} \;\; F_{\mu \nu} = 0.
\label{f12_f34}
\end{eqnarray}
The subscript AH implies traceless anti-hermitian projection.
From the definition in (\ref{topo_charge}), the topological
charge is obtained to be,
\begin{equation}
Q = -\frac{1}{4\pi^2} \sum_x \text{Tr}\, \big[F_{12}\,F_{34}
\big] = -\frac{2L^4}{4\pi^2} \,\sin\left( \frac{2\pi n_1}{L^2} \right)\,
\sin \left( \frac{2\pi n_2}{L^2} \right) \approx -2n_1 n_2, \label{uqtop}
\end{equation}
where the last result is for (moderately) large $L$. For $\tau_8$,
\textit{i.e.} $k=8$, the
above derivation does not hold, but in the large volume limit, it too
gives us $-2n_1 n_2$ as topological charge.

The gauge fields $U_\mu(x)$ thus generated correspond to a given
$Q$ depending on $n_1,\,n_2$. These fields are considered
smooth in the sense that they are not elements of a Markov chain. Such a smooth
configuration may not always give well-separated sign flips in the flow of
eigenvalues of the Dirac operator. Hence `roughening' the
$U_\mu(x)$ by keeping $Q$ approximately invariant has been
suggested \cite{Smit:1986fn,Gattringer:1997c}. The idea is to generate some
$\theta_\mu^k$ from uniformly distributed random numbers $\in [-\delta
\pi, +\delta \pi]$, where $\delta \ll 1$, in order to roughen
the links $U_\mu(x)$ in Eqs. (\ref{u1u2}, \ref{u3u4}),
\begin{equation}
U_\mu^\delta (x) = \exp\left( i\sum_k \theta_\mu^k \,\tau_k \right)
\;\;\rightarrow \;\;  \widetilde{U}_\mu(x) = U_\mu^\delta (x)
U_\mu(x), \label{roughU}
\end{equation}
where $\widetilde{U}_\mu$ are the roughened gauge fields. In all
subsequent calculations with Smit-Vink gauge fields, presented in the
tables and plots in this section, we use such rough gauge fields
$\widetilde{U}_\mu(x)$ generated on $8^3 \times 8$ lattice with $Q=-2$
and $\delta=0.05$.

\subsection{Determination of the near-zero eigenvalues}\label{subsec:determine_eigenvalues}

The computation of zero or near-zero eigenvalues is carried out by using the
Kalkreuter-Simma (KS) algorithm \cite{Kalkreuter:1995mm}. It provides both
eigenvectors and eigenvalues for any hermitian matrix and not just only for
positive definite matrices. We implement the algorithm by suitably
modifying the relevant subroutines in the publicly available MILC
code \cite{MILC}. The hermitian version of the Dirac MDF operator is
\begin{equation}
H(m) = \gamma_5(D + m) \;\;\;\Rightarrow \;\;\; H^2(m) = (D^{\dagger}
+ m)(D + m), \label{mdf_herm_op}
\end{equation}
where $H,\,D$ are either of KW or BC. The eigenvalues of $H(m)$ appear with
both signs; therefore, obtaining the near-zero eigenvalues by
direct computation is not practical even for a modest 4-dimensional
lattice volume. This problem can be sidestepped by using $H^2(m)$
to solve the eigenvalue equation $H^2(m) \vert \psi_j \rangle
= \lambda_j \vert \psi_j \rangle$, where the lowest eigenvalues are always
non-negative and close to zero. However, in general, neither $\vert \psi_j \rangle$
nor $\pm \sqrt{\lambda_j}$ are necessarily the eigenvectors
or eigenvalues of $H(m)$. Nevertheless, to a good approximation, the eigenvectors
of $H(m)$, and hence the near-zero eigenvalues, can be obtained by
following the Rayleigh-Ritz procedure \cite{rayleigh_ritz}. Following
this procedure, once a desired number (typically a few hundred) of the lowest
eigenvalues and their corresponding eigenvectors, say $\vert \psi^\prime \rangle$,
are obtained, a reduced matrix $\mathcal{M}$ of $H(m)$ can be constructed, and
the resulting eigenvalue problem is solved,
\begin{equation}
\mathcal{M}_{xy} = \langle \psi^\prime_x \vert H(m) \vert \psi^\prime_y
\rangle \;\;\;\rightarrow \;\;\; \mathcal{M} \vert \chi_x \rangle = \mu_x
\vert \chi_x \rangle. \label{redmat}
\end{equation}
An approximate eigenvector $\vert \phi \rangle$ of $H(m)$ can
be calculated as
\begin{equation}
\vert \phi_k \rangle = \sum_x \chi_k^x \vert \psi^\prime_x \rangle,
\label{phi_from_psi}
\end{equation}
where $\chi_k^x$ is the $x$-th component of $k$-th normalized eigenvector
of $\mathcal{M}$. This kind of rotation is a feature in the KS algorithm
\cite{Kalkreuter:1995mm}, which is called intermediate
diagonalization, and is used to improve previously computed eigenvectors
of a matrix. The eigenvalue equation for $H(m)$, for lowest lying
eigenvalue, in the subspace of $H^2(m)$ is now
\begin{eqnarray}
& H(m) \vert \phi_k \rangle \approx \mu_k \vert \phi_k \rangle, \label{eval_H} \\
\text{with residue } \;\;\; & r_k = \vert\vert (H(m) - \mu_k )\vert \phi_k \rangle \vert\vert \approx 0. \label{DKresidue}
\end{eqnarray}
\begin{table*}[h]
\centering
\begin{tabular}{ c  c  c |  c  c  c} \hline
        \makecell{Eigenvalue of\\ $H^2_\text{KW}$}
        & \makecell{Eigenvalue of\\ $H_\text{KW}$}
        & $r_k$ 
        & \makecell{Eigenvalue of\\ $H^2_\text{BC}$}
        & \makecell{Eigenvalue of\\ $H_\text{BC}$}
        & $r_k$ 
        \\ \hline
        
        $~~~2.462 \times 10^{-5}~~~$ & $-4.962 \times 10^{-3}~~~$ & $5 \times 10^{-6}~~~$
       &$~~~4.041 \times 10^{-7}~~~$ & $-6.356 \times 10^{-4}~~~$ & $6 \times 10^{-6}$\\
        
        $~~~2.462 \times 10^{-5}~~~$ & $\phantom{-}4.962 \times 10^{-3}~~~$ & $5 \times 10^{-6}~~~$
       &$~~~4.041 \times 10^{-7}~~~$ & $\phantom{-}6.356 \times 10^{-4}~~~$ & $6 \times 10^{-6}$\\

        $~~~4.808 \times 10^{-5}~~~$ & $-6.934 \times 10^{-3}~~~$ & $6 \times 10^{-6}~~~$
       &$~~~1.820 \times 10^{-5}~~~$ & $-4.266 \times 10^{-3}~~~$ & $5 \times 10^{-6}$\\

        $~~~4.808 \times 10^{-5}~~~$ & $\phantom{-}6.934 \times 10^{-3}~~~$ & $6 \times 10^{-6}~~~$
       &$~~~1.820 \times 10^{-5}~~~$ & $\phantom{-}4.266 \times 10^{-3}~~~$ & $5 \times 10^{-6}$\\

        $~~~2.420 \times 10^{-3}~~~$ & $-4.919 \times 10^{-2}~~~$ & $7 \times 10^{-6}~~~$
       &$~~~2.955 \times 10^{-3}~~~$ & $-5.436 \times 10^{-2}~~~$ & $4 \times 10^{-6}$\\

        $~~~2.420 \times 10^{-3}~~~$ & $\phantom{-}4.919 \times 10^{-2}~~~$ & $7 \times 10^{-6}~~~$
       &$~~~2.955 \times 10^{-3}~~~$ & $\phantom{-}5.436 \times 10^{-2}~~~$ & $4 \times 10^{-6}$\\

        $~~~2.468 \times 10^{-3}~~~$ & $-4.968 \times 10^{-2}~~~$ & $7 \times 10^{-6}~~~$
       &$~~~3.038 \times 10^{-3}~~~$ & $-5.512 \times 10^{-2}~~~$ & $5 \times 10^{-6}$\\

        $~~~2.468 \times 10^{-3}~~~$ & $\phantom{-}4.968 \times 10^{-2}~~~$ & $7 \times 10^{-6}~~~$
       &$~~~3.038 \times 10^{-3}~~~$ & $\phantom{-}5.512 \times 10^{-2}~~~$ & $5 \times 10^{-6}$\\

        $~~~2.905 \times 10^{-3}~~~$ & $-5.389 \times 10^{-2}~~~$ & $6 \times 10^{-6}~~~$
       &$~~~3.186 \times 10^{-3}~~~$ & $-5.644 \times 10^{-2}~~~$ & $6 \times 10^{-6}$\\
     \hline 
\end{tabular}
\caption{For $m=0$, we tabulate eigenvalues
of $H^2$ and $H$ with residues $r_k$ as in Eq. (\ref{DKresidue})
for $Q=-2$ and $\delta=0.05$ background $8^3 \times 8$
Smit-Vink lattice.}
\label{table:diff_eigenvalue_m0}
\end{table*}

\begin{table*}[h]
\centering
\begin{tabular}{ c  c  c |  c  c  c} \hline
         \makecell{Eigenvalue of\\ $H^2_\text{KW}$}
         & \makecell{Eigenvalue of\\ $H_\text{KW}$}
         & $r_k$
         & \makecell{Eigenvalue of\\ $H^2_\text{BC}$}
         & \makecell{Eigenvalue of\\ $H_\text{BC}$}
         & $r_k$
         \\ \hline

        $~~~1.442 \times 10^{-2}~~~$ & $-1.201 \times 10^{-1}~~~$ & $5 \times 10^{-6}~~~$
       &$~~~1.440 \times 10^{-2}~~~$ & $-1.200 \times 10^{-1}~~~$ & $6 \times 10^{-6}$\\
        
        $~~~1.442 \times 10^{-2}~~~$ & $\phantom{-}1.201 \times 10^{-1}~~~$ & $5 \times 10^{-6}~~~$
       &$~~~1.440 \times 10^{-2}~~~$ & $\phantom{-}1.200 \times 10^{-1}~~~$ & $6 \times 10^{-6}$\\

        $~~~1.445 \times 10^{-2}~~~$ & $-1.202 \times 10^{-1}~~~$ & $6 \times 10^{-6}~~~$
       &$~~~1.442 \times 10^{-2}~~~$ & $-1.201 \times 10^{-1}~~~$ & $4 \times 10^{-6}$\\

        $~~~1.445 \times 10^{-2}~~~$ & $\phantom{-}1.202 \times 10^{-1}~~~$ & $6 \times 10^{-6}~~~$
       &$~~~1.442 \times 10^{-2}~~~$ & $\phantom{-}1.201 \times 10^{-1}~~~$ & $5 \times 10^{-6}$\\

        $~~~1.682 \times 10^{-2}~~~$ & $-1.297 \times 10^{-1}~~~$ & $7 \times 10^{-6}~~~$
       &$~~~1.735 \times 10^{-2}~~~$ & $-1.317 \times 10^{-1}~~~$ & $3 \times 10^{-6}$\\

        $~~~1.682 \times 10^{-2}~~~$ & $\phantom{-}1.297 \times 10^{-1}~~~$ & $7 \times 10^{-6}~~~$
       &$~~~1.735 \times 10^{-2}~~~$ & $\phantom{-}1.317 \times 10^{-1}~~~$ & $5 \times 10^{-6}$\\

        $~~~1.687 \times 10^{-2}~~~$ & $-1.299 \times 10^{-1}~~~$ & $6 \times 10^{-6}~~~$
       &$~~~1.744 \times 10^{-2}~~~$ & $-1.321 \times 10^{-1}~~~$ & $3 \times 10^{-6}$\\

        $~~~1.687 \times 10^{-2}~~~$ & $\phantom{-}1.299 \times 10^{-1}~~~$ & $7 \times 10^{-6}~~~$
       &$~~~1.744 \times 10^{-2}~~~$ & $\phantom{-}1.321 \times 10^{-1}~~~$ & $3 \times 10^{-6}$\\

        $~~~1.730 \times 10^{-2}~~~$ & $-1.315 \times 10^{-1}~~~$ & $6 \times 10^{-6}~~~$
       &$~~~1.759 \times 10^{-2}~~~$ & $-1.326 \times 10^{-1}~~~$ & $4 \times 10^{-6}$\\
        \hline
\end{tabular}
\caption{Same as Table. \ref{table:diff_eigenvalue_m0}
but for some small, non-zero mass $m=0.12$.}
\label{table:diff_eigenvalue_m0p12}
\end{table*}

\noindent
The $\ket{\phi_k}$ is the approximate eigenvector with approximate eigenvalue $\mu_k$,
\textit{i.e.}, $r_k \approx 0$. For `good' eigenvalues and hence `good' eigenvectors,
$r_k$ should be small.\footnote{While using the Rayleigh-Ritz procedure \cite{Kishore:2025fxt},
we encounter some spurious eigenvalues of $H$ at the top end of the near-zero eigenmodes.
These originate from errors in determining the corresponding eigenvalues and
eigenvectors of $H^2$. We discard a few such high-lying eigenmodes, thereby eliminating
the spurious eigenmodes and improving the accuracy of the eigenmodes of $H$.}
In Table \ref{table:diff_eigenvalue_m0} and \ref{table:diff_eigenvalue_m0p12}
we present a representative sample of residues for $m=0$ and $m=0.12$, respectively. Eigenvalues
of $H^2$ are calculated with a numerical tolerance of $\mathcal{O}(10^{-6})$.

\subsection{Flavored mass and Index theorem}\label{subsec:index_flav_mass}

\begin{figure}[htb]
\makebox[\textwidth]{\includegraphics[width=0.45\textwidth]{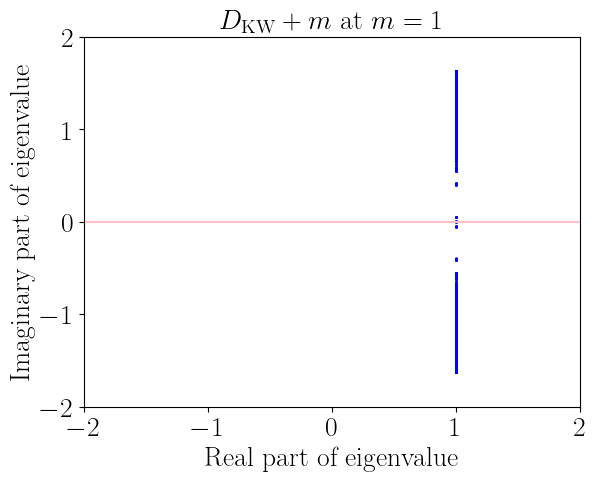} \hspace{0.1in}
\includegraphics[width=0.45\textwidth]{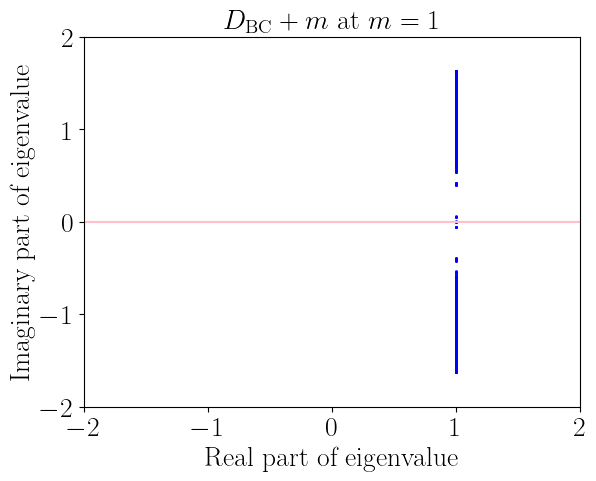}}
\vspace{-0.2in}
\caption{Complex eigenvalue plots of KW and BC fermions for $Q=-2$ and $\delta=0.05$ background $8^3 \times 8$
Smit-Vink lattice with mass parameter $m=1$ for 5000 eigenvalues. The left and
right panel correspond to $(D_\text{KW}+m)$ and $(D_\text{BC}+m)$.}\label{fig:complex_dirac_eigval}
\end{figure}

In the Fig. \ref{fig:normal_mass_specflow}, we were unable to detect the index
using spectral flow, despite using gauge fields with a specific non-zero integer
value of $Q$. For MDF with degenerate doublers,
the index gets cancelled between the pairs and hence the flow of eigenvalue with such
mass term $m (1 \otimes 1)$ does not result in net crossings at $m \approx 0$.
The degeneracy in masses for $\pm$ chiral states is made apparent from the plot of the complex
eigenvalues (Eq. (\ref{evDg5D})) of the operator $\big(D+ m(1 \otimes 1)\big)$
in Fig. \ref{fig:complex_dirac_eigval}.

\begin{figure}[htb]
\makebox[\textwidth]{\includegraphics[width=0.45\textwidth]{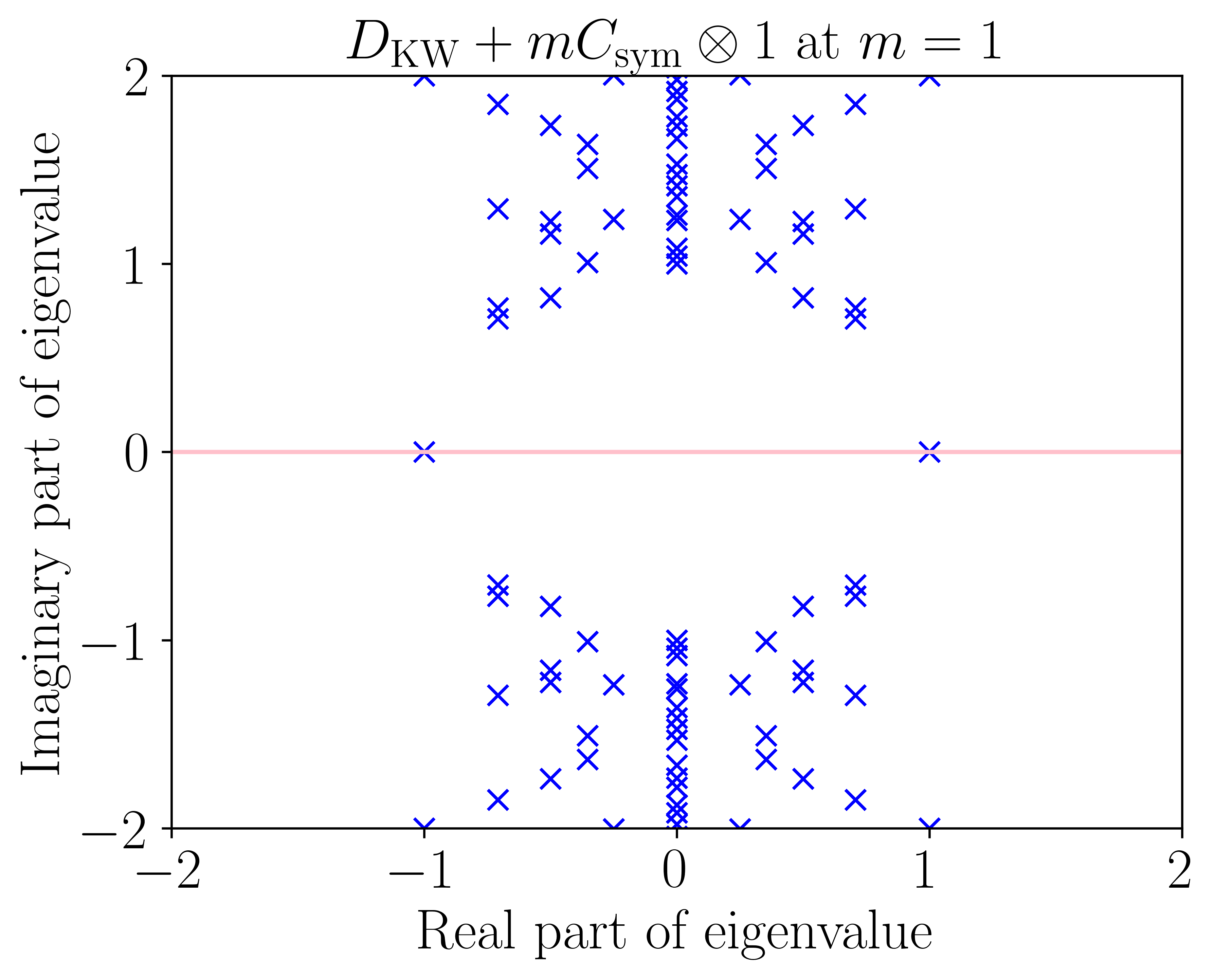} \hspace{0.1in}
\includegraphics[width=0.45\textwidth]{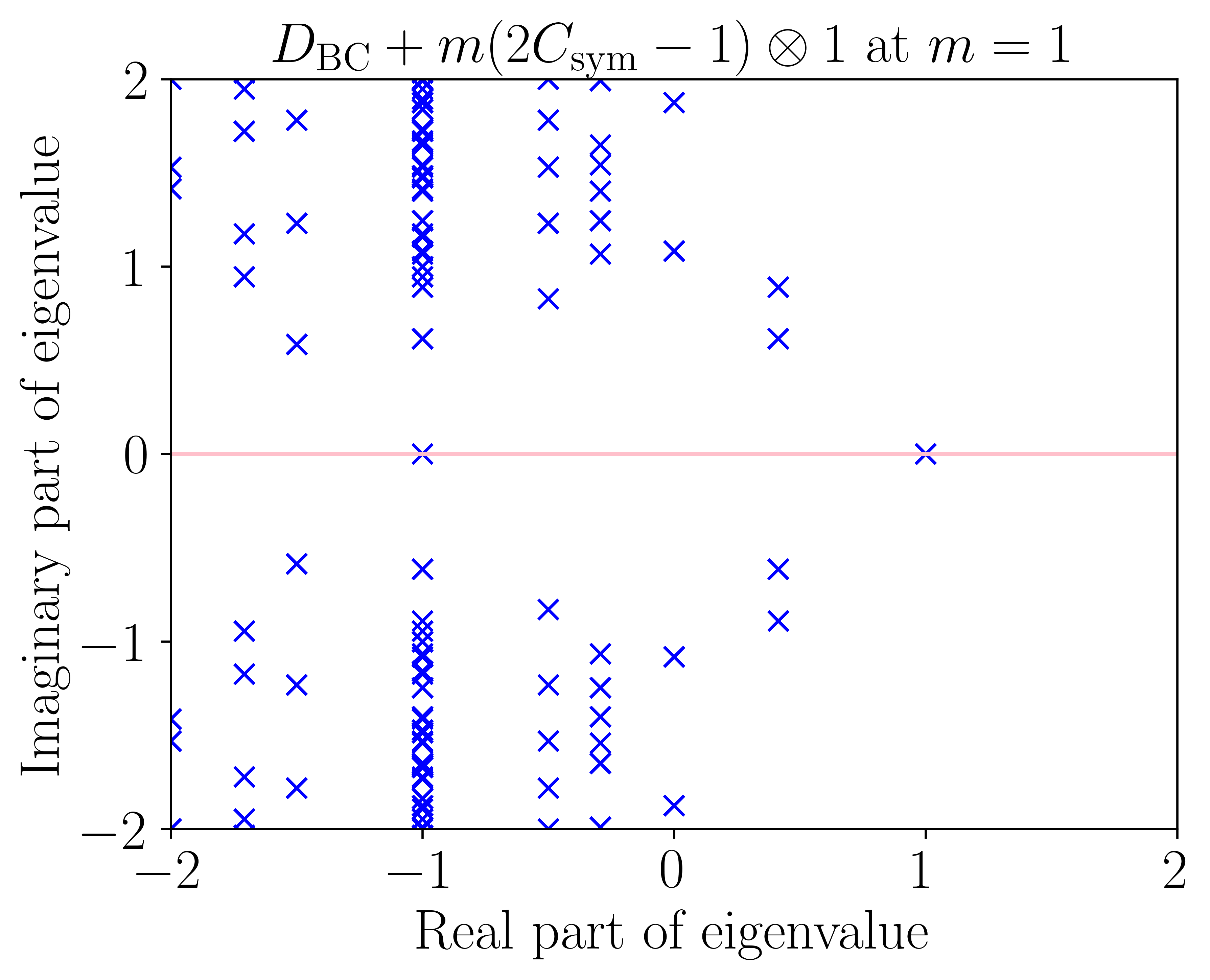}}
\vspace{-0.2in}
\caption{Complex eigenvalue plots of KW and BC fermions for free theory
on $8^3\times8$ lattice with mass parameter $m=1$. The left and
right panel correspond to $(D_{\text{KW}}+mC_{\text{sym}}\otimes 1)$
and $(D_{\text{BC}}+m(2C_{\text{sym}}-1)\otimes 1)$ operators,
respectively.}
\label{fig:complex_dirac_csym_eigval_free_th}
\end{figure}

\begin{figure}[htb]
\makebox[\textwidth]{\includegraphics[width=0.45\textwidth]{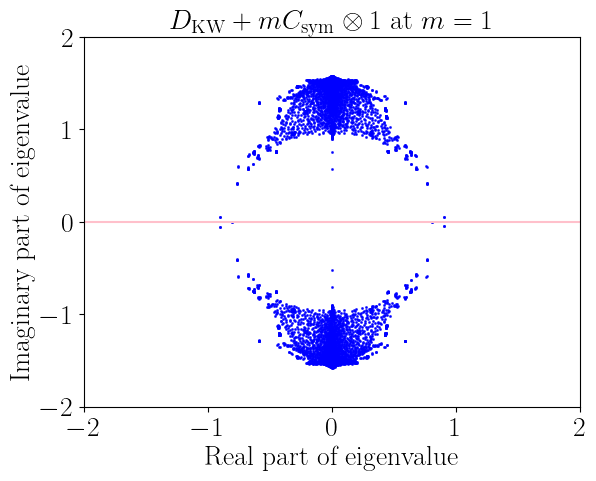} \hspace{0.1in}
\includegraphics[width=0.45\textwidth]{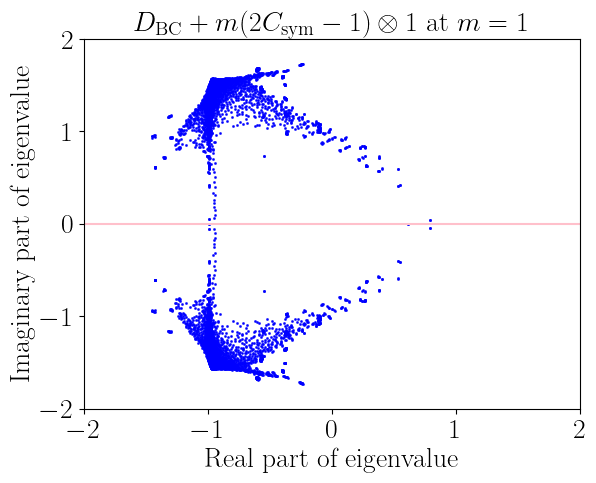}}
\vspace{-0.2in}
\caption{Same plot as Fig. \ref{fig:complex_dirac_csym_eigval_free_th} but
with $Q=-2$ and $\delta=0.05$ background $8^3 \times 8$ Smit-Vink lattice.}
\label{fig:complex_dirac_csym_eigval}
\end{figure}

\noindent
The problem of degeneracy in real eigenvalues can be resolved by lifting
the degeneracy of the masses of $\pm$ chiral states by introducing a doubler
or taste dependent so-called `flavored mass term', $m C_{\text{flav}}\otimes1$
\cite{Creutz:2010bm,Durr:2022mnz}
\begin{eqnarray}
&& C_{\text{flav}}^{\text{KW}} = C_{\text{sym}} \;\;\;\;\text{and}
\;\;\; C_{\text{flav}}^{\text{BC}} = 2C_{\text{sym}} -1 \label{cflav}, \\
\text{where} \;\;\; && C_{\text{sym}} = \frac{1}{4!}
\sum_{\substack{\text{perm}}} C_1 C_2 C_3 C_4, \label{csym_op} \\
\text{and}\;\;\; && C_{\mu}(x,y) \ket{\psi(y)} = \frac{1}{2} \big[
U_{\mu}(x) \delta_{x+\hat{\mu},y} + U^{\dagger}_{\mu}(x-\hat{\mu})
\delta_{x-\hat{\mu},y} \big] \ket{\psi(y)}. \label{csym_ex}
\end{eqnarray}
The eigenvalues of the flavored MDF operators $(D+mC_{\text{flav}}
\otimes 1)$ are shown in Fig. \ref{fig:complex_dirac_csym_eigval_free_th}
for free theory. The separation of the zero eigenvalues along the
real axis at $\pm m$ for KW, and appropriately shifted for BC, shows
removal of the degeneracy discussed above. When the background gauge field
with a particular topological charge is turned on, we observe similar
splittings of the low-lying states as shown in Fig. \ref{fig:complex_dirac_csym_eigval}.

Next, we calculate the eigenvalues of $(D + mC_\text{flav})$ with Smit-Vink gauge field.
Since $(D + mC_\text{flav})$ is a non-hermitian and non-normal operator, its eigenvalues
are complex, and left- and right-eigenvectors are not related by the adjoint operation
\cite{Hip:2001mh}. To obtain its eigenvalues, we construct a reduced matrix for $(D + mC_\text{flav})$
using the eigenvectors of hermitian $(D^\dagger + mC_\text{flav})(D + mC_\text{flav})$, and
compute the eigenvalues of this reduced matrix. We found many `spurious' eigenvalues in this
set but as the size of the reduced matrix is increased, these spurious eigenvalues disappear.

\begin{figure}[htb]
\makebox[\textwidth]{\includegraphics[width=0.45\textwidth]{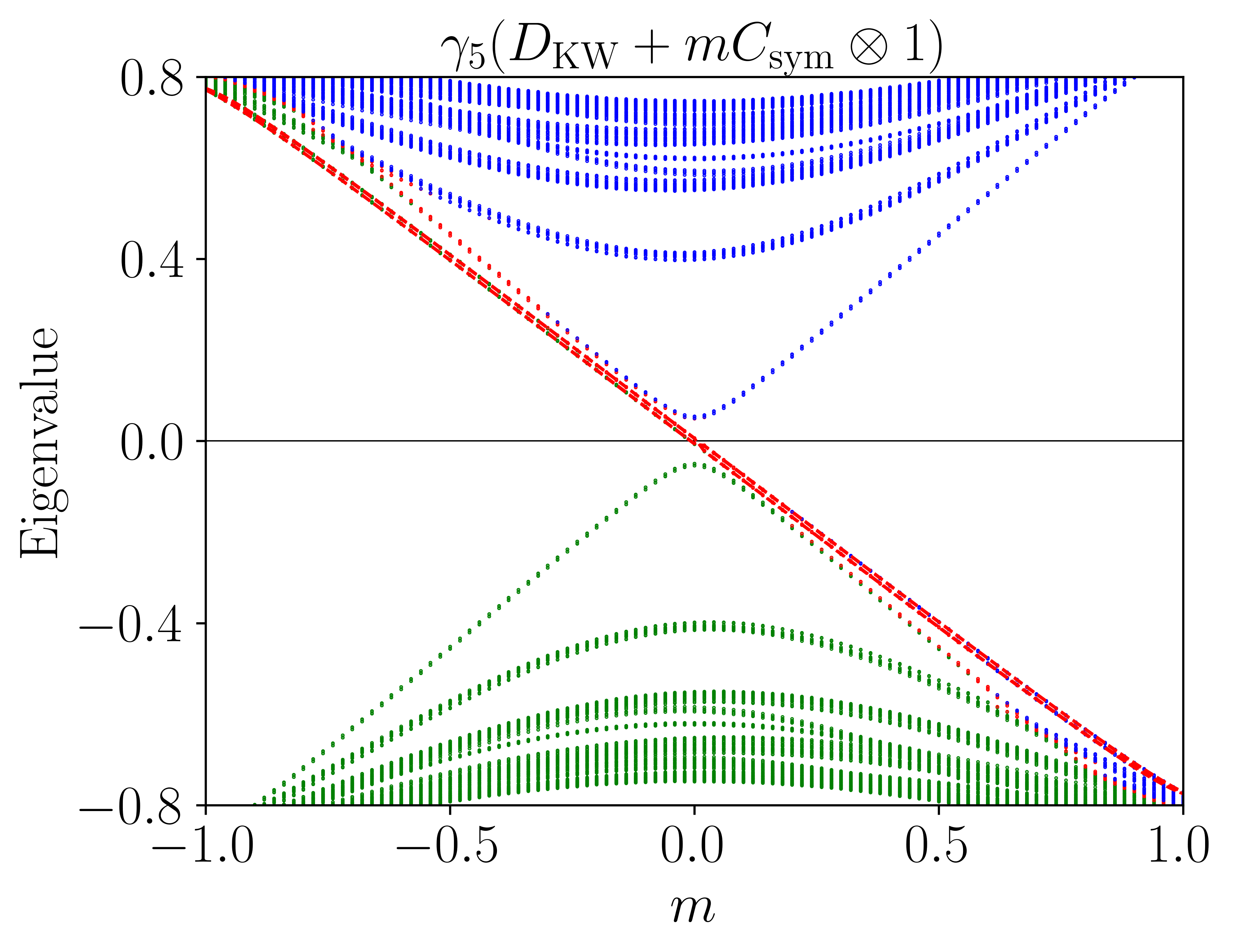} \hspace{0.1in}
\includegraphics[width=0.45\textwidth]{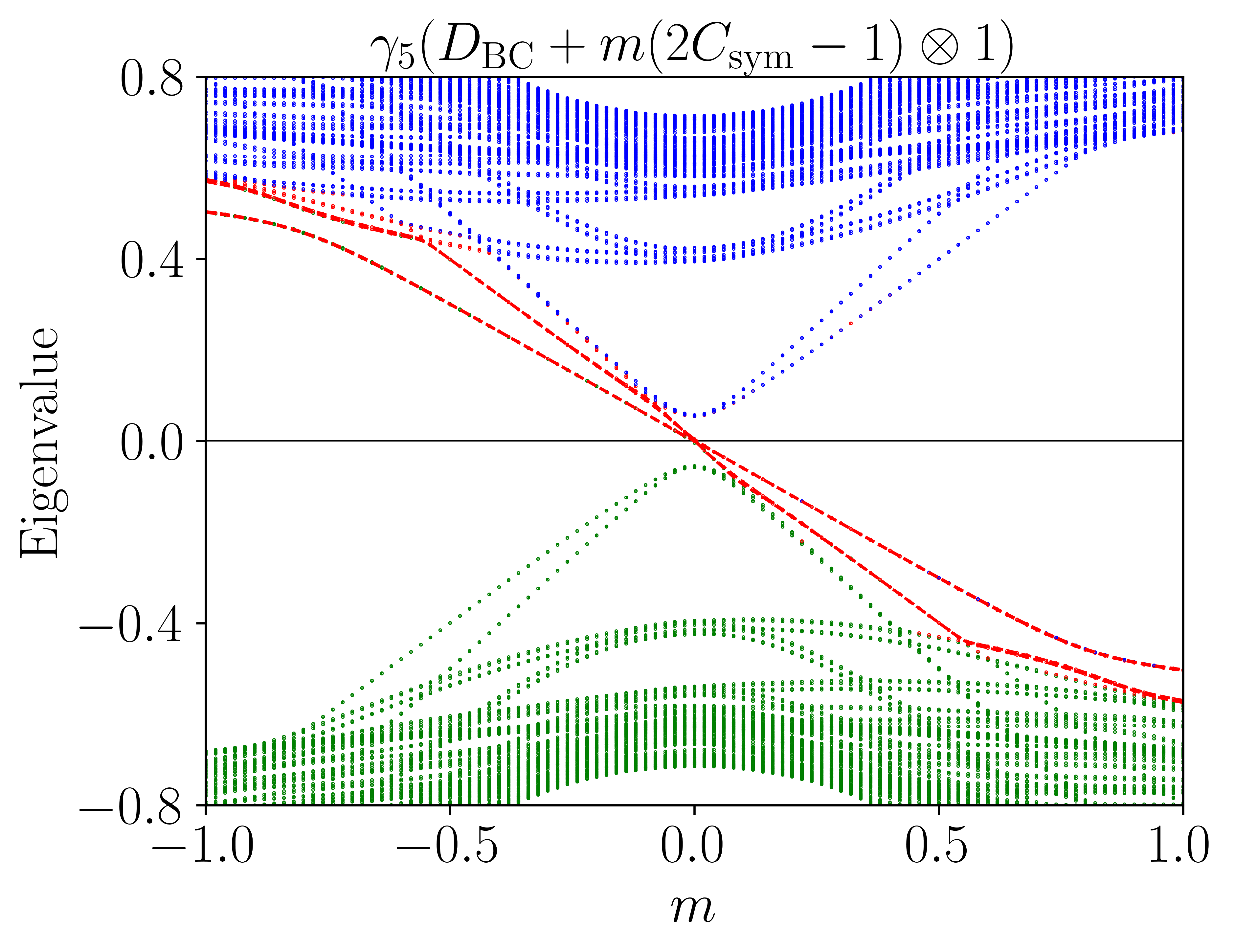}}
\vspace{-0.2in}
\caption{Spectral flow of the eigenvalues of $\mathcal{H}$
with respect to $m$ for $Q=-2$ and $\delta=0.05$ background Smit-Vink lattice.
 Four nearly overlapping lines (red in color) cross the zero eigenvalue line at $m=0$.
The color coding is intended to guide the eye.}
\label{fig:flavored_mass_specflow}
\end{figure}

When the simple mass-term is replaced by $mC_{\text{flav}}$ mass for
both KW and BC fermions, the spectral flow in Fig. \ref{fig:flavored_mass_specflow}
exhibits the exact number of crossings as required by the expression in Eq. (\ref{qtopmdf}).
We have two lines crossing near $m=0$ with negative slopes and each being two-fold degenerate,
for both KW and BC fermions.
The flow shown in Fig. \ref{fig:flavored_mass_specflow} corresponds to the
eigenvalues of the associated Hermitian MDF operators
\begin{eqnarray}
\mathcal{H}_\text{KW}(m) &=& \gamma_5 \big(D_{\text{KW}} + m C_{
\text{sym}}\otimes 1 \big), \label{hkw5} \\
\mathcal{H}_\text{BC}(m) &=& \gamma_5 \big(D_{\text{BC}} + m
(2C_{\text{sym}} - 1) \otimes1 \big). \label{hbc5}
\end{eqnarray}
In this case, the index theorem acquires an additional factor of 2 arising
from two non-degenerate masses of the doubler species, depending on their
$\pm$ chirality \cite{Creutz:2010bm}
\begin{equation}
\text{index}(D) = 2Q.
\label{qtopmdf}
\end{equation}
To obtain the near-zero modes of $\mathcal{H}$, we proceed as before
to first find the lowest-lying eigenvalues of the operator $\mathcal{H}^2$.
Followed by constructing a reduced matrix $\mathcal{M}$ as in Eq.
(\ref{redmat}) and its eigenstates, whose linear combination serves as
basis states (rotational basis) of $\mathcal{H}$. The result is
tabulated below in Table.  \ref{table:diff_eigenvalue_mathcal_m0p12_l88q2delta0p05}.

\begin{table*}[h]
\centering
\begin{tabular}{ c  c  c  |  c  c  c} \hline
         \makecell{Eigenvalue of\\ $\mathcal{H}^2_\text{KW}$}
         & \makecell{Eigenvalue of\\ $\mathcal{H}_\text{KW}$}
         & $r_k$
         & \makecell{Eigenvalue of\\ $\mathcal{H}^2_\text{BC}$}
         & \makecell{Eigenvalue of\\ $\mathcal{H}_\text{BC}$}
         & $r_k$
        \\ \hline
		
        $~~~8.055 \times 10^{-3}~~~$ & $ -8.975 \times 10^{-2}~~~$ & $5 \times 10^{-7}~~~$
       &$~~~5.120 \times 10^{-3}~~~$ & $-7.155 \times 10^{-2}~~~$ & $1 \times 10^{-6}$\\
        
        $~~~8.435 \times 10^{-3}~~~$ & $-9.184 \times 10^{-2}~~~$ & $5 \times 10^{-7}~~~$
       &$~~~5.229 \times 10^{-3}~~~$ & $-7.231 \times 10^{-2}~~~$ & $1 \times 10^{-6}$\\

        $~~~1.034 \times 10^{-2}~~~$ & $-1.017 \times 10^{-1}~~~$ & $4 \times 10^{-7}~~~$
       &$~~~1.102 \times 10^{-2}~~~$ & $-1.050 \times 10^{-1}~~~$ & $1 \times 10^{-6}$\\

        $~~~1.065 \times 10^{-2}~~~$ & $-1.032 \times 10^{-1}~~~$ & $4 \times 10^{-7}~~~$
       &$~~~1.202 \times 10^{-2}~~~$ & $-1.096 \times 10^{-1}~~~$ & $1 \times 10^{-6}$\\

        $~~~1.380 \times 10^{-2}~~~$ & $\phantom{-}1.175 \times 10^{-1}~~~$ & $4\times 10^{-7}~~~$
       &$~~~1.202 \times 10^{-2}~~~$ & $\phantom{-}1.096 \times 10^{-1}~~~$ & $1 \times 10^{-6}$\\

        $~~~1.388 \times 10^{-2}~~~$ & $\phantom{-}1.178 \times 10^{-1}~~~$ & $6 \times 10^{-7}~~~$
       &$~~~1.210 \times 10^{-2}~~~$ & $\phantom{-}1.100 \times 10^{-1}~~~$ & $2 \times 10^{-6}$\\

        $~~~1.388 \times 10^{-2}~~~$ & $-1.178 \times 10^{-1}~~~$ & $5 \times 10^{-7}~~~$
       &$~~~1.440 \times 10^{-2}~~~$ & $-1.200 \times 10^{-1}~~~$ & $1 \times 10^{-6}$\\

        $~~~1.404 \times 10^{-2}~~~$ & $-1.185 \times 10^{-1}~~~$ & $5 \times 10^{-7}~~~$
       &$~~~1.561 \times 10^{-2}~~~$ & $-1.250 \times 10^{-1}~~~$ & $1 \times 10^{-6}$\\

        $~~~1.497 \times 10^{-2}~~~$ & $\phantom{-}1.223 \times 10^{-1}~~~$ & $3 \times 10^{-7}~~~$
       &$~~~1.732 \times 10^{-2}~~~$ & $\phantom{-}1.316 \times 10^{-1}~~~$ & $1 \times 10^{-6}$\\
        \hline
\end{tabular}
\caption{For $m=0.12$, we tabulate eigenvalues
of $\mathcal{H}^2$ and $\mathcal{H}$ with residues $r_k$
for $Q=-2$ and $\delta=0.05$ background $8^3 \times 8$ Smit-Vink lattice.}
\label{table:diff_eigenvalue_mathcal_m0p12_l88q2delta0p05}
\end{table*}

To determine the chiral states of the eigenvectors $|\psi
\rangle$ of the MDF operator $D$ in Eq. (\ref{evDg5D}), we observe that
both $|\psi \rangle$ and $\gamma_5 |\psi \rangle$ are eigenvectors with
eigenvalues $\pm i\lambda$. For $\lambda = 0$, the corresponding eigenvector
can also be chosen as a simultaneous eigenvector of $\gamma_5$
(see Eq. (\ref{eq:definite_chirality})).
In general, when we calculate the eigenvalues of
$D$, then we do not impose the condition on eigenvectors of $D$ with
zero eigenvalues to be simultaneous eigenvectors of $\gamma_5$.
Consequently, arbitrary linear combinations of these chiral states are also eigenvectors of $D$
with $\lambda=0$. In such cases, we get $\langle \psi | \gamma_5 | \psi
\rangle \approx 0$. A brief discussion on this aspect is provided in
Appendix \ref{appendix_B} (see Eq. (\ref{eq:arb_g5_expec})). Our numerical results
yield $\langle \gamma_5 \rangle \approx 0$, implying that the
zero-mode eigenvectors in our setup are not simultaneous eigenvectors of $D$ and $\gamma_5$.
Therefore, we cannot determine the chiralities of the zero eigenmodes
with $\gamma_5$ operator. Hence, modified chirality operators $\mathcal{X}$
are proposed \cite{Durr:2022mnz}
\begin{eqnarray}
\mathcal{X}_{\text{KW}}=C_{\text{sym}}\otimes\gamma_5 \;\;\;
\text{and} \;\;\; \mathcal{X}_{\text{BC}}=(2C_{\text{sym}}-1)\otimes
\gamma_5. \label{xchi}
\end{eqnarray}
In Table \ref{table:chirality_l88q2delta0p05}, we list the lowest-lying
eigenvalues and their corresponding chiralities, computed using the
eigenvectors of $H(m=0)$ \cite{Blum:2001qg} rather than $D$. This choice is
motivated by the poor residues obtained when attempting to construct
the eigenvectors of $D$ from $H^2(m=0)$ for various lattice sizes. 
Using $H(m=0)$ is justified for two reasons: (i) for zero-modes $(\lambda=0)$,
the eigenvectors of $D$ and $H(m=0)$ are identical (see Eq. (\ref{eq:D_H_eigvec})),
and (ii) for $\lambda \neq 0$, the expectation value $\langle \psi | \gamma_{5} | \psi \rangle$
vanishes for either operator (see Eqs. (\ref{eq:g5_chirality},
\ref{eq:non_zero_eigenmode_H_chirality})). While $\langle \gamma_5 \rangle \approx 0$
throughout, the modified chirality operator $\mathcal{X}$ correctly identifies
the chiralities for zero eigenmodes for any $Q$, yielding an index of
$2Q$, in agreement with the index theorem. Notably, since $C_\text{flav}$
does not commute with $D$ in presence of gauge fields \cite{Creutz:2010bm}, the
modified chirality $\mathcal{X}$ can never attain exact $\pm1$ for chiral states.

\begin{figure}[htb]
\includegraphics[width=0.5\textwidth]{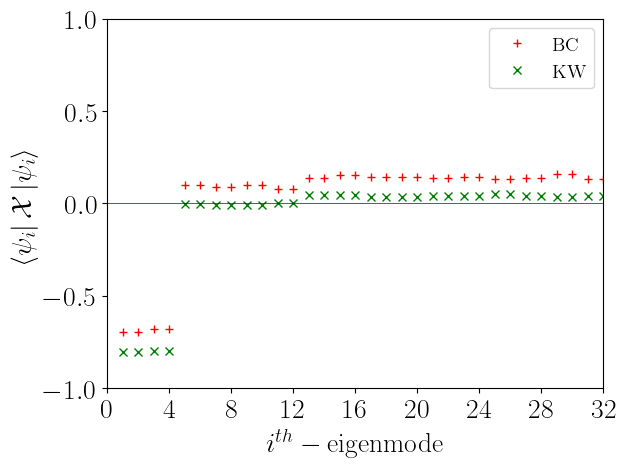}
\vspace{-0.2in}
\caption{The modified chirality $\mathcal{X}$ for $Q=-2$ and $\delta=0.05$ background $8^3 \times 8$ Smit-Vink lattice.
For clarity, $\langle \mathcal{X}_\text{BC} \rangle$ is shifted by $+0.1$.}
\label{fig:chirality_vs_index_l88q2delta0p05}
\end{figure}

\begin{figure}[H]
\makebox[\textwidth]{
\includegraphics[width=0.45\textwidth]{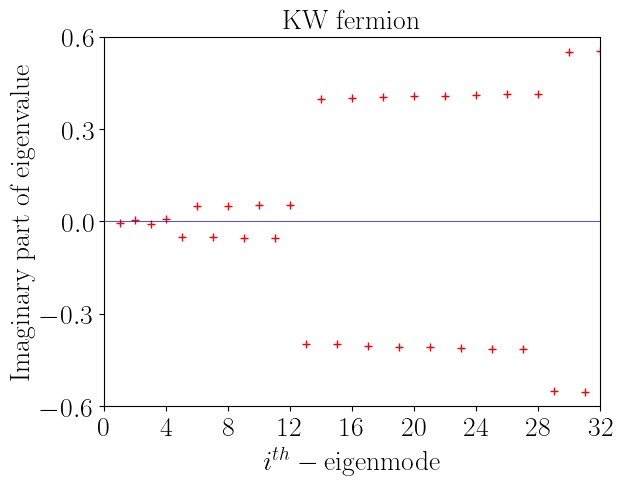} \hspace{0.1in}
\includegraphics[width=0.45\textwidth]{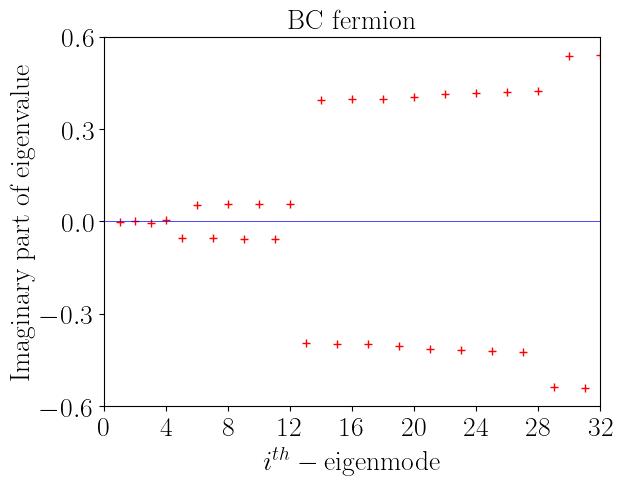}
}
\vspace{-0.2in}
\caption{Non-zero eigenvalues form clusters of doublets, while the
number of zero eigenvalues
depends on the background topology. The results are for $Q=-2$ and $\delta=0.05$
background $8^3 \times 8$ Smit-Vink lattice.}
\label{fig:eigenvalue_vs_index_l88q2delta0p05}
\end{figure}

\begin{table}[H]
\centering
\begin{tabular}{ c  c | c  c} \hline
         \makecell{Eigenvalue of $H_\text{KW}(m=0)$}
         & $\bra{\psi_i}\mathcal{X}_{\text{KW}}\ket{\psi_i}$
         & \makecell{Eigenvalue of $H_\text{BC}(m=0)$}
         & $\bra{\psi_i}\mathcal{X}_{\text{BC}}\ket{\psi_i}$
         \\ \hline
      $-4.962 \times 10^{-3}$  & $-8.037 \times 10^{-1}~~~$
      & $-6.356 \times 10^{-4}$  & $-7.935 \times 10^{-1}$\\
		
      $\phantom{-}4.962 \times 10^{-3}$  & $-8.037 \times 10^{-1}~~~$ 
      & $\phantom{-}6.356 \times 10^{-4}$  & $-7.935 \times 10^{-1}$\\
		
      $-6.934 \times 10^{-3}$  & $-7.995 \times 10^{-1}~~~$
      & $-4.266 \times 10^{-3}$  & $-7.771 \times 10^{-1}$\\
          
	   $\phantom{-}6.934 \times 10^{-3}$  & $-7.995 \times 10^{-1}~~~$
      & $\phantom{-}4.266 \times 10^{-3}$  & $-7.771 \times 10^{-1}$\\
          
	   $-4.919 \times 10^{-2}$  & $-9.096 \times 10^{-4}~~~$
      & $-5.436 \times 10^{-2}$  & $-1.892 \times 10^{-3}$\\
          
	   $\phantom{-}4.919 \times 10^{-2}$  & $-9.096 \times 10^{-4}~~~$
      & $\phantom{-}5.436 \times 10^{-2}$  & $-1.892 \times 10^{-3}$\\
          
	   $-4.968 \times 10^{-2}$  & $-6.392 \times 10^{-3}~~~$
      & $-5.512 \times 10^{-2}$  & $-1.106 \times 10^{-3}$\\ \hline
\end{tabular}
\caption{Eigenvalues and modified chirality $\mathcal{X}$
using eigenvectors of $H(m=0)$ for $Q=-2$ and $\delta=0.05$ background $8^3 \times 8$ Smit-Vink lattice.}
\label{table:chirality_l88q2delta0p05} 
\end{table}
\noindent
In Fig. \ref{fig:chirality_vs_index_l88q2delta0p05}, we
plotted the expectation of modified chirality operator $\mathcal{X}$ against various
eigenmodes, among which we found 4 of them with chirality $-1$ (approximately).
The imaginary part of the eigenvalues of $D$
are shown in Fig. \ref{fig:eigenvalue_vs_index_l88q2delta0p05}.
We observe that non-zero eigenvalues of $D$ form clusters of doublets because of the doubler species.

\begin{figure}[htb]
\includegraphics[width=0.5\textwidth]{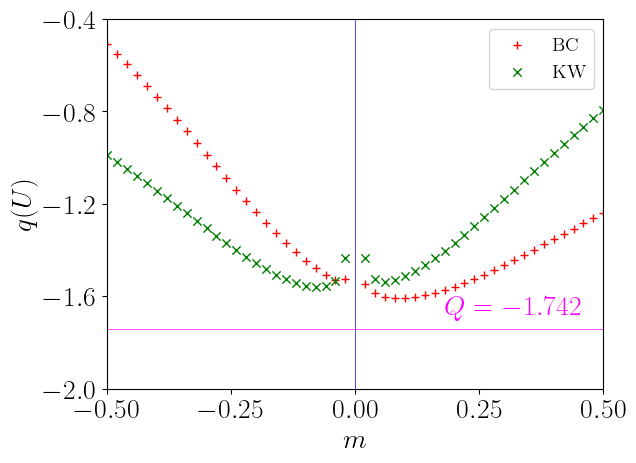}
\vspace{-0.2in}
\caption{Variation of fermionic topological charge with $m$ for $Q=-2$ and $\delta=0.05$ background $8^3 \times
8$ Smit-Vink Lattice. $Q$ is measured to be $-1.742$.}
\label{fig:fermi_topo_charge}
\end{figure}

An alternative way to verify the index theorem is using the fermionic
topological charge \cite{Smit:1986fn,Smit:1987fq,Kerler:1999pw,Durr:2022mnz}.
In the continuum on a background field $A_{\mu}$ with topological charge
$Q \in \mathbb{Z}$, it is given by 
\begin{equation}
    q(A) = \lim_{m \to 0} m \text{Tr}\big[\gamma_5(D+m)^{-1}\big].
    \label{q_ferm_continuum}
\end{equation}
For MDF in the presence of background gauge fields $U_\mu$, this expression becomes
\begin{equation}
q(U) = \lim_{m \to 0} \frac{m}{2} \,\text{Tr}\big[
(C_\text{flav} \otimes \gamma_5)\, (D + m)^{-1}\big]. \label{fermtopo}
\end{equation}
In Fig. \ref{fig:fermi_topo_charge}, if we neglect a small pole like structure near $m = 0$ then we can read
the $q(U)$ to be approximately $-1.6$. It deviates slightly from the $Q=-1.742$ measured using
Eq. (\ref{uqtop}) with the improved field strength tensor $F^I_{\mu\nu}$ given in Eq. (\ref{fs_def}).
This may be attributed to the roughening of $Q=-2$ Smit-Vink ﬁeld, which is performed in a way that
approximately preserves the original $Q$. We observe that in Fig. \ref{fig:fermi_topo_charge}
an apparent divergence of $q(U)$ appears around $m=0$, which vanishes
for larger lattice volume. Its appearance could be due to our use of a rather small
lattice volume of $8^3 \times 8$.

\section{Index theorem on dynamical QCD lattices}
\label{sec:index_under_general_gauge_field}

In the previous section, we showed that the index theorem holds for MDF in 4-dimension
on Smit-Vink type lattices. In this section, we attempt to explore how the
theorem holds up on QCD lattices. For this purpose, we use MILC asqtad
$N_f=2+1$ lattices \cite{Bernard:2001av}. We choose coarse $16^3 \times 48$
lattices with $10/g_0^2 = 6.572$ and $a \approx 0.15$ fm followed by smoothing
steps, which is called ``cooling".
Smoothing steps have been central to the study of topology and instantons in QCD
\cite{Bonnet:2001rc}, as the topological charge of gauge fields is highly
sensitive to these procedures. In particular, smoothing/cooling significantly
improves the reliability of identifying topological charges. This is essential
to accurately establish the universality of the expression in Eq. (\ref{qtopmdf}).
For this work, we utilize the cooling algorithm described in Ref. \cite{Bonnet:2001rc};
however, unlike that approach, we do not perform any smearing steps. To proceed
further, we reproduce some of its expressions that are relevant for our case.

\begin{figure}[htb]
\includegraphics[width=0.5\textwidth]{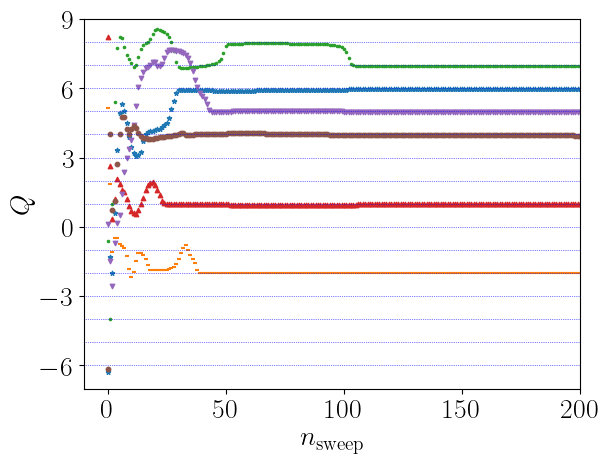}
\vspace{-0.2in}
\caption{Variation of $Q$ as a function of cooling sweeps
$n_{\text{sweep}}$ on six configurations of MILC $16^3 \times 48$ lattice.}
\label{fig:cooling}
\end{figure}

The cooling algorithm essentially constructs improved staples to define an
``improved" topological charge. The improved staples $\Sigma_{\mu \nu}^I(x)$
consist of a linear combination of square and rectangular staples in $\mu - \nu$ plane,
\begin{eqnarray}
\Sigma_{\mu \nu}^I(x) &=& \frac{5}{3} \Sigma_{\mu \nu}(x) +
\frac{1}{12u_0^2} \Sigma_{\mu \nu}^R(x), \label{improved_staple} \\
\text{where} \;\;\; \Sigma_{\mu \nu}(x) &=& U_{\nu}(x+\hat{\mu})\,
U^\dagger_{\mu}(x+\hat{\nu})\, U^\dagger_{\nu}(x) + U^\dagger_{\nu}
(x+\hat{\mu}-\hat{\nu})\, U^\dagger_{\mu}(x-\hat{\nu})\, U_{\nu}
(x-\hat{\nu}), \label{plaq_staple} \\
\text{and} \;\;\; \Sigma_{\mu \nu}^R(x) &=& U_{\mu}(x+\hat{\mu})\,
U_{\nu}(x+2\hat{\mu})\,U^\dagger_{\mu}(x+\hat{\mu}+\hat{\nu}) \,
U^\dagger_{\mu}(x+\hat{\nu})\, U^\dagger_{\nu}(x) + \nonumber \\
&& U_{\nu}(x+\hat{\mu})\, U_{\nu}(x+\hat{\mu} + \hat{\nu})\,
 U^\dagger_{\mu}(x+2\hat{\nu})\, U^\dagger_{\nu}(x+\hat{\nu}) \,
U^\dagger_{\nu}(x) + \nonumber \\
&& U_{\nu}(x+\hat{\mu})\, U^\dagger_{\hat{\mu}}(x+\hat{\nu})\, 
U^\dagger_{\mu}(x-\hat{\mu}+\hat{\nu})\, U^\dagger_{\nu}(x-\hat{\mu})
\,U_{\mu}(x-\hat{\mu}) + \nonumber \\
&& U^\dagger_{\nu}(x+\hat{\mu}-\hat{\nu})\, U^\dagger_{\mu}(x-\hat{\nu})
\,U^\dagger_{\mu}(x-\hat{\mu}-\hat{\nu}) \,U_{\nu}(x-\hat{\mu}-\hat{\nu})
\,U_{\mu}(x-\hat{\mu}) + \nonumber \\
&& U^\dagger_{\nu}(x+\hat{\mu}-\hat{\nu})\, U^\dagger_{\nu}(x+\hat{\mu}
-2\hat{\nu})\, U^\dagger_{\mu}(x-2\hat{\nu})\, U_{\nu}(x-2\hat{\nu})
\,U_{\nu}(x-\hat{\nu}) + \nonumber \\
&& U_{\mu}(x+\hat{\mu})\, U^\dagger_{\nu}(x+2\hat{\mu}-\hat{\nu})\,
U^\dagger_{\mu}(x+\hat{\mu}-\hat{\nu})\, U^\dagger_{\mu}(x-\hat{\nu})
\,U_{\nu}(x-\hat{\nu}), \label{rect_staple}
\end{eqnarray}
where $u_0$ is the tadpole factor that gets updated after each iteration through the lattice.
The action of cooling results in replacing the original links $U_\mu(x)$ by the links that maximizes
\begin{equation}
\text{max} \,\, \text{Re} \,\, \text{Tr}\Big( U_{\mu}(x) \sum_{
\substack{\nu \\ \nu \neq \mu}}  \Sigma_{\mu \nu}^I(x)  \Big).
\label{smoothlink}
\end{equation}
The improved links define the $\mathcal{O}(a^2)$ improved field strength tensor $F_{\mu \nu}^I(x)$ which
is used for topological charge to ensure (nearly) integer value of $Q$ defined in Eq. (\ref{topo_charge}).
\begin{eqnarray}
F_{\mu \nu}^I(x) &=& \frac{5}{3} \Omega_{\mu \nu}^{(1,1)}(x) -
\frac{1}{3} \Omega_{\mu \nu}^{(1,2)}(x), \label{fs_def} \\
\text{where} \;\; \Omega_{\mu \nu}^{(1,1)}(x) &=& \frac{1}{2i} \big(
C_{\mu\nu}^{(1,1)}(x) - C_{\mu \nu}^{(1,1)\dagger} \big)_\text{AH},
\label{omega11} \\
\text{and} \;\;\Omega_{\mu \nu}^{(1,2)}(x) &=& \frac{1}{2i} \big(
C_{\mu \nu}^{(1,2)}(x) - C_{\mu \nu}^{(1,2)\dagger} \big)_\text{AH}.
\label{omega12}
\end{eqnarray}
The subscript AH implies traceless anti-hermitian projection. The $C_{\mu\nu}$'s are
clover leaf arrangements of $1\times 1$ square and $2 \times 1$ rectangular plaquettes,
\begin{eqnarray}
P_{\mu \nu}^{(1\times1)}(x) &=& U_{\mu}(x)\, U_{\nu}(x+\hat{\mu})\,
U_{\mu}^{\dagger}(x+\hat{\nu})\, U_{\nu}^{\dagger}(x), \label{plaq} \\
P_{\mu \nu}^{(1\times2)}(x) &=& U_{\mu}(x)\, U_{\nu}(x+\hat{\mu})\,
U_{\nu}(x+\hat{\mu}+\hat{\nu})\, U_{\mu}^{\dagger}(x+2\hat{\nu})\,
U_{\nu}^{\dagger}(x+\hat{\nu})\, U_{\nu}^{\dagger}(x),
\label{rect_plaq12} \\
P_{\mu \nu}^{(2\times1)}(x) &=& U_{\mu}(x)\, U_{\mu}(x+\hat{\mu})\,
U_{\nu}(x+2\hat{\mu})\, U_{\mu}^{\dagger}(x+\hat{\mu}+\hat{\nu})\,
U_{\mu}^{\dagger}(x+\hat{\nu})\, U_{\nu}^{\dagger}(x), \label{rect_plaq21} \\
C_{\mu \nu}^{(1,1)}(x) &=& \frac{1}{4} \big(P_{\mu \nu}^{(1\times1)}(x)
+ P_{-\mu \nu}^{(1\times1)}(x) + P_{\mu -\nu}^{(1\times1)}(x) +
P_{-\mu -\nu}^{(1\times1)}(x) \big), \label{clover_plaq} \\
C_{\mu \nu}^{(1,2)}(x) &=& \frac{1}{8} \big( P_{\mu \nu}^{(2\times1)}(x)
+ P_{-\mu \nu}^{(2\times1)}(x) + P_{\mu -\nu}^{(2\times1)}(x) +
P_{-\mu -\nu}^{(2\times1)}(x) \nonumber \\
  && + P_{\mu \nu}^{(1\times2)}(x) + P_{-\mu \nu}^{(1\times2)}(x) +
P_{\mu -\nu}^{(1\times2)}(x) + P_{-\mu -\nu}^{(1\times2)}(x) \big).
\label{clover_rect_plaq}
\end{eqnarray}
The topological charge $Q$, now defined using $F^I_{\mu \nu}$, is plotted as a
function of the number of cooling sweeps $n_{\text{sweep}}$ in Fig. \ref{fig:cooling}.
Cooling reduces the fluctuations in $Q$ and leads to stable plateaus, typically
after 60 sweeps, at near-integer values. This evolution of $Q$ is found consistent
with Ref. \cite{Bonnet:2001rc}.

\begin{figure}[htb]
\includegraphics[width=0.5\textwidth]{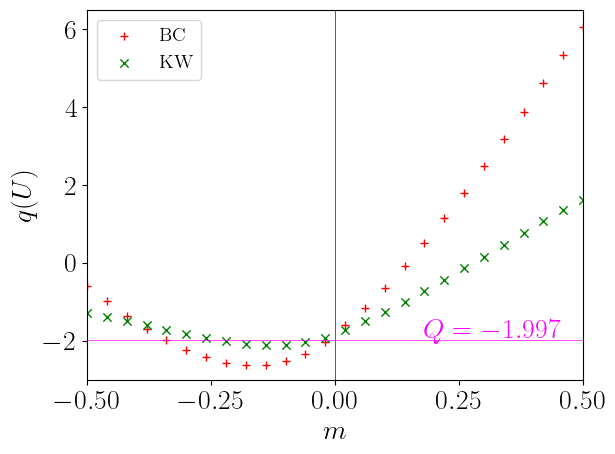}
\vspace{-0.2in}
\caption{Variation of fermionic topological charge with $m$ for $Q=-1.997$ background
$16^3 \times 48$ MILC lattice.}
\label{fig:fermi_topo_charge_qm2}
\end{figure}

To remain consistent with the previous case of $Q=-2$ for Smit-Vink lattice, we select the MILC
configuration with $Q=-1.997$, which is closest to $-2$, for the spectral flow and subsequent analysis.
Before analyzing
the spectral flow, we cross-check the $Q$ value for the chosen configuration by calculating the
fermionic topological charge $q$, as defined in Eq. (\ref{fermtopo}).
As in the previous section, we examine the mass dependence of $q$ and verify whether
$q(m \rightarrow0) \approx Q$. This behaviour is illustrated in Fig. \ref{fig:fermi_topo_charge_qm2}
for $Q \approx -2$. The same procedure is applied to a few other configurations with different $Q$,
for example see Fig. \ref{fig:fermi_topo_charge_q3} for $Q \approx -3$.

\begin{figure}[H]
\makebox[\textwidth]{
\includegraphics[width=0.45\textwidth]{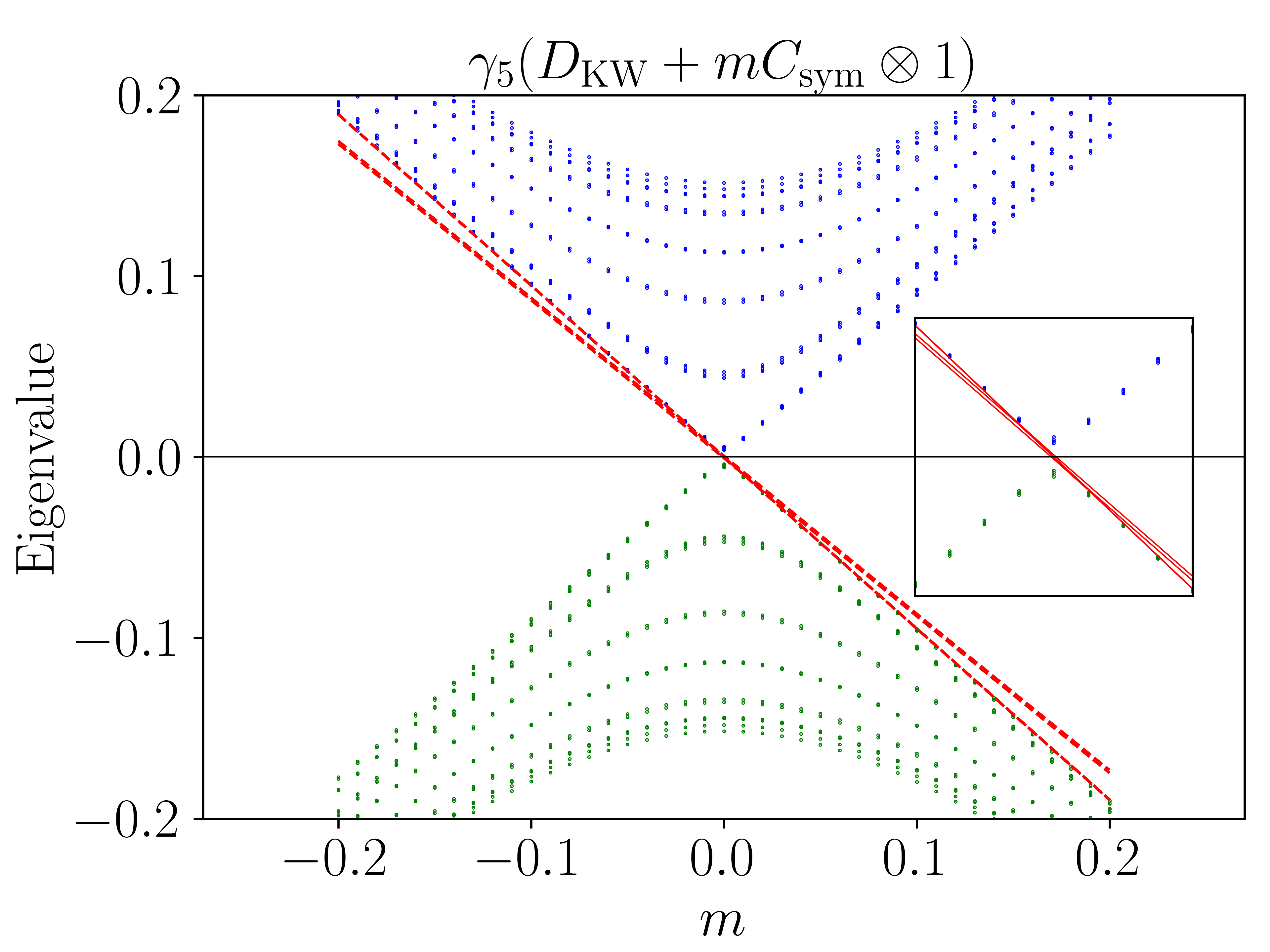} \hspace{0.1in}
\includegraphics[width=0.45\textwidth]{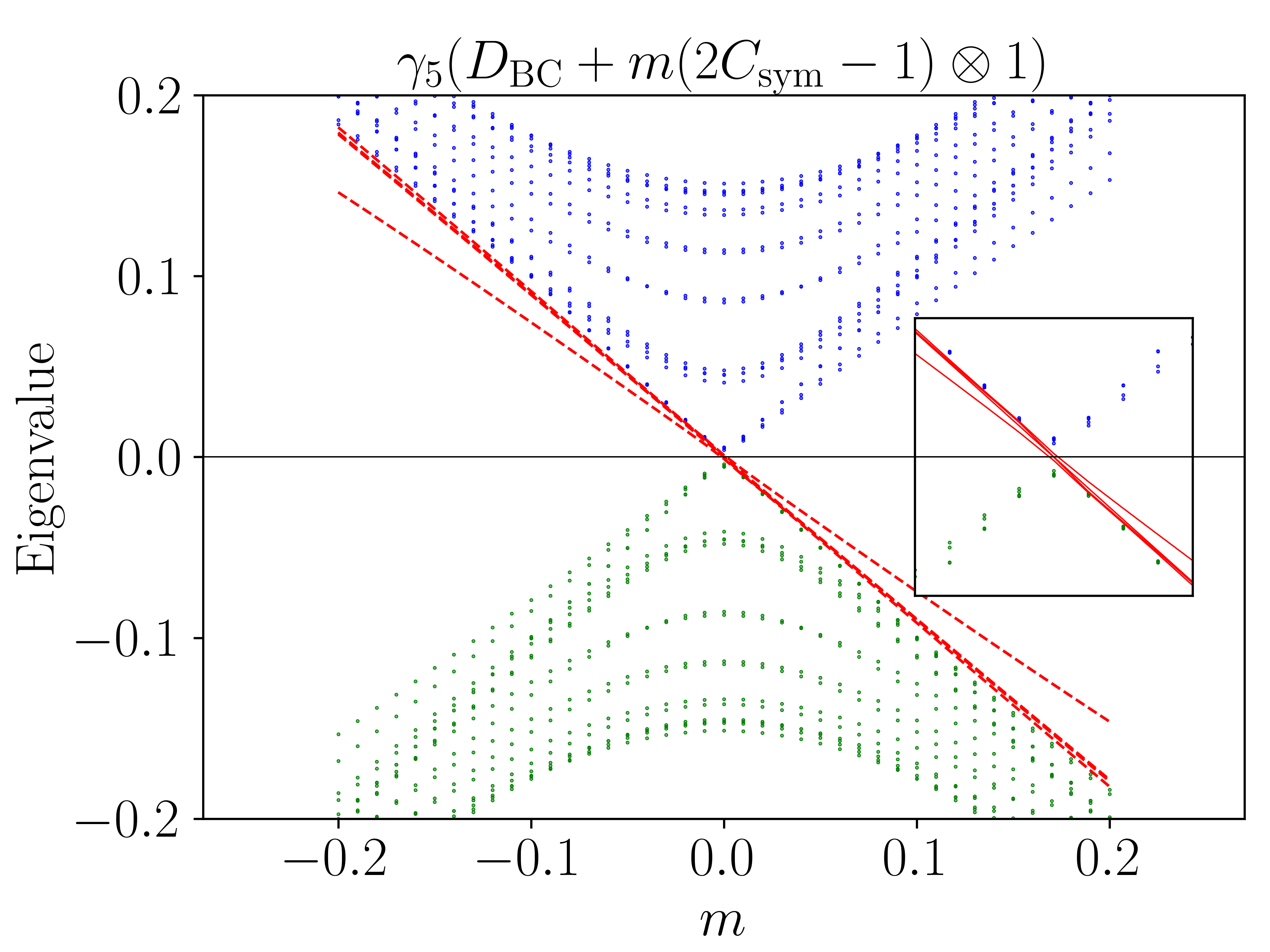}
}
\vspace{-0.2in}
\caption{Spectral flow of the eigenvalues of $\mathcal{H}$ with respect to $m$ for $Q=-1.997$ background
$16^3 \times 48$ MILC lattice. In the insets, the x-and y-axis ranges are from $-0.04$ to $0.04$.
Total four nearly overlapping lines (red in color) cross the zero eigenvalue line at $m=0$.
The color coding is intended to guide the eye.}  \label{fig:specflow_milc_config_qm2}
\end{figure}

For the MILC configuration with $Q \approx -2$, we present in Fig. \ref{fig:specflow_milc_config_qm2}
the spectral flow of the near-zero eigenvalues of the hermitian MDF Dirac operator with a flavor-dependent mass
term. The net number of eigenvalue crossings is consistent with index $=-4 \approx 2Q$.

\begin{figure}[htb]
\includegraphics[width=0.5\textwidth]{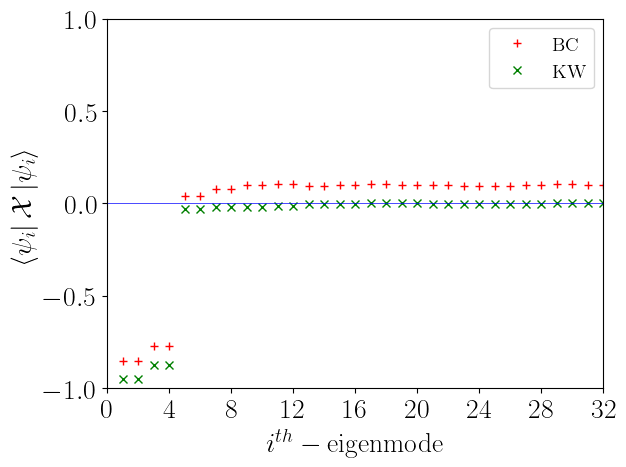}
\vspace{-0.2in}
\caption{The modified chirality $\mathcal{X}$ for
$Q=-1.997$ background $16^3 \times 48$ MILC lattice.
For clarity, $\langle \mathcal{X}_\text{BC} \rangle$ is shifted by $+0.1$.}
\label{fig:chirality_vs_index_qm2}
\end{figure}

Unsurprisingly, with the MILC lattices too we find $\langle \gamma_5 \rangle \approx 0$ and
consequently $\gamma_5$ fails to distinguish the chiralities of the low-lying modes. This
necessitates the use of modified chirality operator $\mathcal{X}$ defined in Eq. (\ref{xchi}).
A selection of near-zero eigenvalues, in increasing order, of $H(m=0) \equiv \mathcal{H}(m=0)$ and
its chirality determined from the $\langle \mathcal{X} \rangle$ are given in Table 
\ref{table:chirality_qm2}. The corresponding plot is given in Fig. \ref{fig:chirality_vs_index_qm2}.

\begin{table}[htb]
\centering
\begin{tabular}{c  c | c  c} \hline
    \makecell{Eigenvalue of $H_\text{KW}(m=0)$}
         & $\bra{\psi_i}\mathcal{X}_{\text{KW}}\ket{\psi_i}$
         & \makecell{Eigenvalue of $H_\text{BC}(m=0)$}
         & $\bra{\psi_i}\mathcal{X}_{\text{BC}}\ket{\psi_i}$
         \\ \hline
	$-4.467 \times 10^{-5}$  & $-9.490 \times 10^{-1}~~~$
    & $-4.079 \times 10^{-5}$  & $-9.504 \times 10^{-1}$\\
          
	$\phantom{-}4.467 \times 10^{-5}$  & $-9.490 \times 10^{-1}~~~$
    & $\phantom{-}4.079 \times 10^{-5}$  & $-9.504 \times 10^{-1}$\\
          
	$-6.052 \times 10^{-4}$  & $-8.762 \times 10^{-1}~~~$
    & $-1.011 \times 10^{-3}$  & $-8.738 \times 10^{-1}$\\
          
	$\phantom{-}6.052 \times 10^{-4}$  & $-8.762 \times 10^{-1}~~~$
    & $\phantom{-}1.011 \times 10^{-3}$  & $-8.738 \times 10^{-1}$\\
          
	$-3.882 \times 10^{-3}$  & $-3.170 \times 10^{-2}~~~$
    & $-3.942 \times 10^{-3}$  & $-6.118 \times 10^{-2}$\\
          
	$\phantom{-}3.882 \times 10^{-3}$  & $-3.170 \times 10^{-2}~~~$
    & $\phantom{-}3.942 \times 10^{-3}$  & $-6.118 \times 10^{-2}$\\
          
	$-4.509 \times 10^{-3}$  & $-1.666 \times 10^{-2}~~~$
    & $-4.940 \times 10^{-3}$  & $-1.978 \times 10^{-2}$\\ \hline
\end{tabular}
\caption{Eigenvalues and modified chirality $\mathcal{X}$ using eigenvectors of
$H(m=0)$ for $Q = -1.997$ background $16^3 \times 48$ MILC lattice.}
\label{table:chirality_qm2} 
\end{table}

\section{Summary and Conclusion}\label{sec:conclusion}
The major advantage of MDF is that they describe two flavors of fermions
on a lattice without explicitly breaking the chiral symmetry. This makes
MDF a strong candidate for lattice simulations of chiral fermions using an ultralocal action.
So far, the chiral properties of MDF, particularly KW and BC fermions, have
been studied primarily in two space-time dimensions. A natural next step is
to investigate them in four space-time dimensions before considering MDF for
realistic QCD calculations. In this work, we have shown
that in four space-time dimensions both KW and BC fermions satisfy the
Atiyah-Singer index theorem on the lattice in the presence of integer
topological charge. The spectral flow of the eigenvalues
of the hermitian Dirac operator is used to determine the index on the lattice.
We consider two types of gauge backgrounds: the Smit-Vink gauge fields with
fixed topological charge and the MILC $N_f = 2 + 1$ asqtad lattices. To improve
the signals, the Smit-Vink configurations are appropriately roughened, while the
MILC lattices are cooled prior to the measurement of the topological
charge and the spectral flow.

The two doublers (or tastes) of MDF with degenerate masses cancel the contributions of opposite
chiral states, resulting in a vanishing index despite a non-zero topological charge background.
To resolve this, a flavor-dependent mass term, determined by the chiral charges, is introduced.
This mass term assigns positive and negative masses to different flavors. The hermitian MDF Dirac
operator with the flavor-dependent mass term then yields an index consistent with the index theorem,
as observed from the spectral flow of its eigenvalues.
The theorem is further verified by measuring the fermionic topological charge. To distinguish
the chirality of the zero and non-zero eigenmodes, a corresponding modification of $\gamma_5$ is
required. The resulting flavor-dependent chirality operator successfully separates the chiralities.

\section*{Acknowledgments}

\noindent The numerical computations have been performed on the HPC and PARAM Sanganak facilities at IIT Kanpur, funded by DST and IIT Kanpur, as well as on the Proton cluster at NISER, Bhubaneswar. A.K. thanks Stephan Dürr for valuable discussions during LATTICE2024 and subsequent email correspondence, and Taro Kimura for useful email communications. He is especially grateful to Johannes Weber for extensive discussions that ultimately led to the removal of spurious eigenvalues of the hermitian Dirac operator. He also thanks Radhamadhab, Sachin, Aritra, Indrajit, and Akash for insightful discussions on eigenvalues and for fostering a stimulating research environment in the lab at NISER. He acknowledges Diptarko for assistance with Python code for eigenvalue computations, and Abhishek, Manisha, and Dipak at IIT Kanpur for helpful discussions that enhanced his understanding of topology and numerical techniques for eigenvalue calculations.

\appendix

\section{ Some results for background MILC lattice with $Q=2.995$}\label{appendix_A}

Here we present some results  for both KW and BC fermions with MILC lattice background gauge field for topological charge $Q=2.995$. All the calculations are performed on  a $16^3\times 48$ lattice. 

\begin{figure}[H]
\makebox[\textwidth]{
\includegraphics[width=0.45\textwidth]{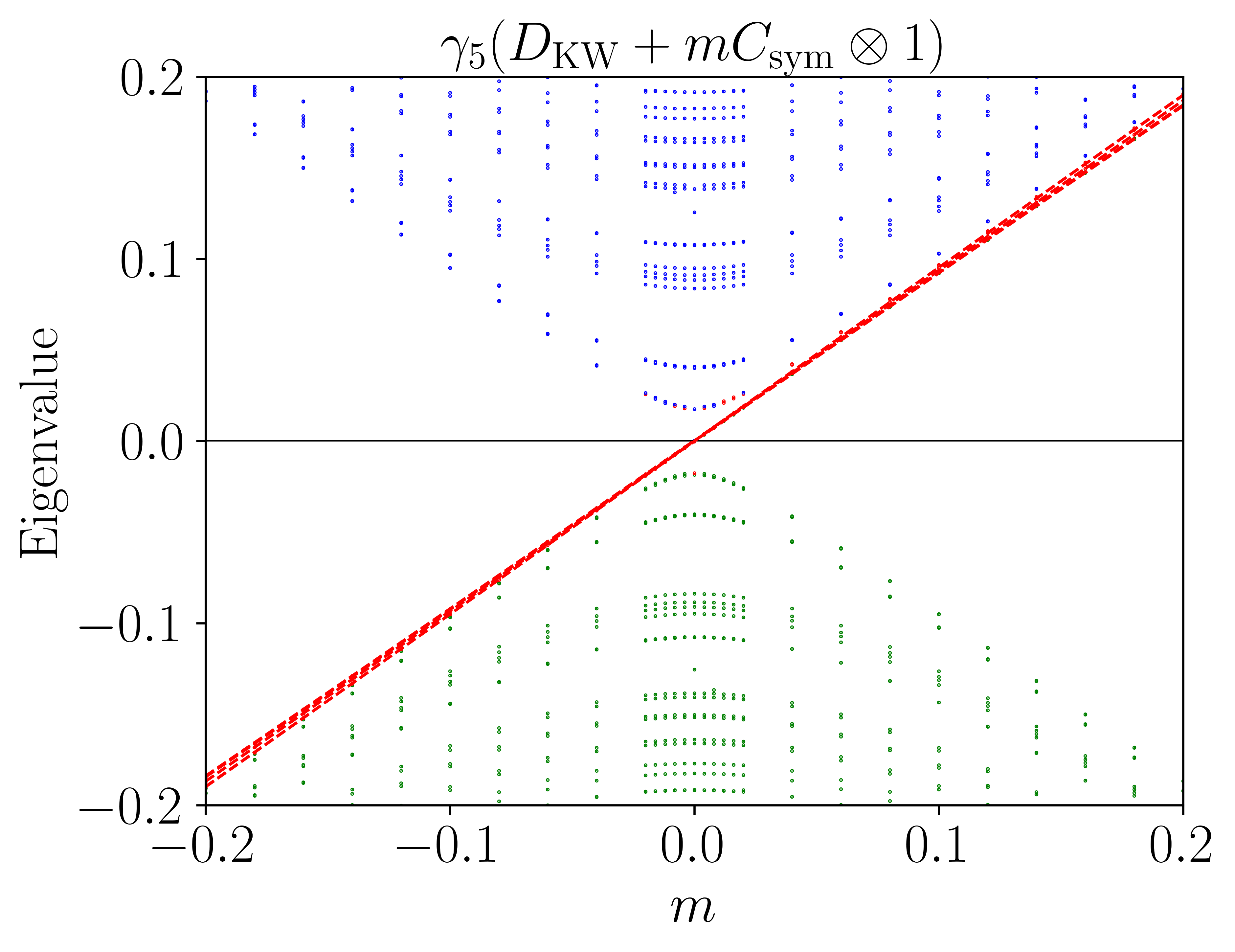} \hspace{0.1in}
\includegraphics[width=0.45\textwidth]{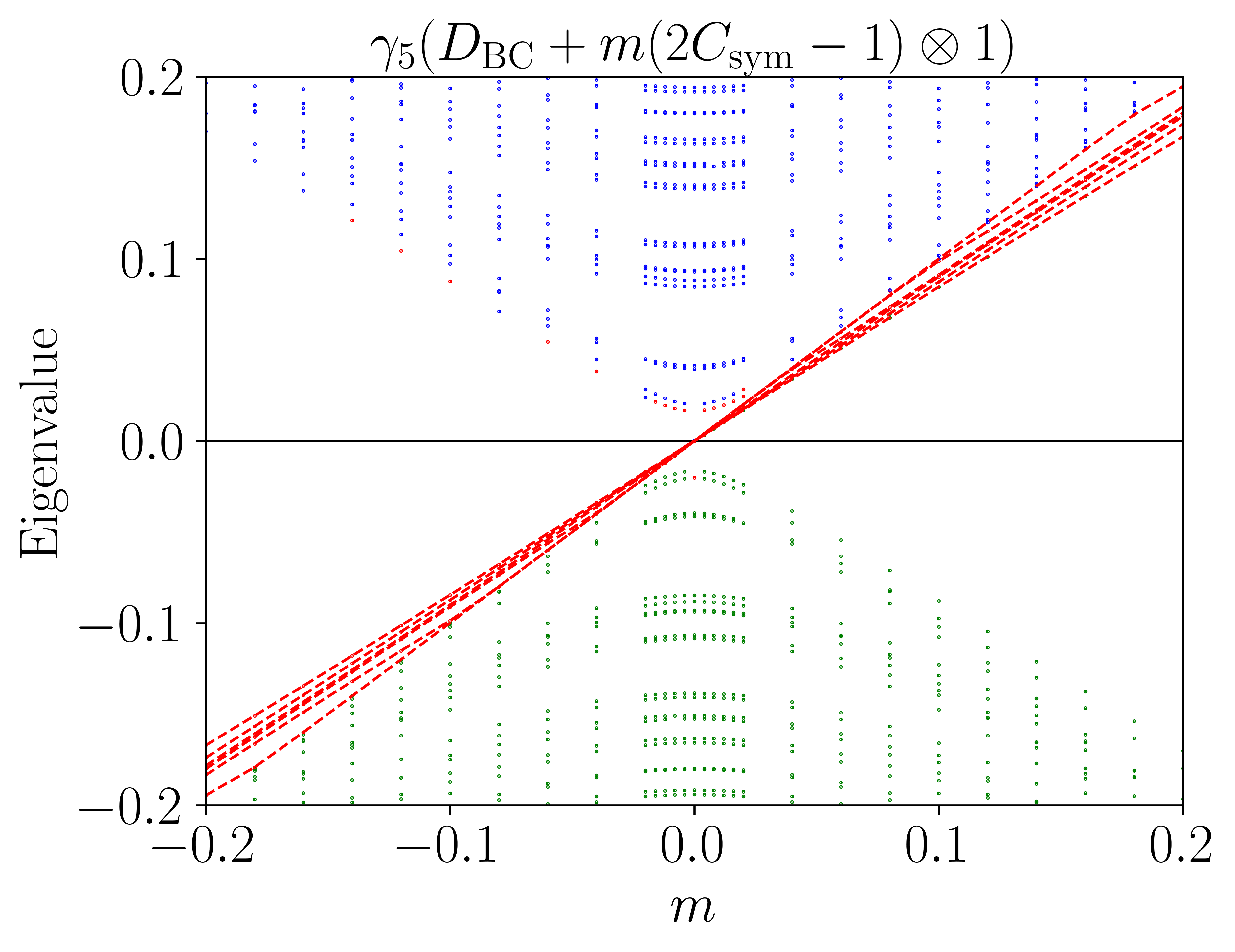}
}
\vspace{-0.2in}
\caption{Spectral flow of the eigenvalues of $\mathcal{H}$ with respect to $m$
for $Q=2.995$ background $16^33 \times 48$ MILC lattice. Total six nearly overlapping lines (red in color) cross the zero eigenvalue line at $m=0$.
The color coding is intended to guide the eye.}
\label{fig:specflow_milc_config_q3}
\end{figure}

\begin{table}[H]
	\centering
	\begin{tabular}{ c  c | c  c} \hline
           \makecell{Eigenvalue of $H_\text{KW}(m=0)$}
         & $\bra{\psi_i}\mathcal{X}_{\text{KW}}\ket{\psi_i}$
         & \makecell{Eigenvalue of $H_\text{BC}(m=0)$}
         & $\bra{\psi_i}\mathcal{X}_{\text{BC}}\ket{\psi_i}$
         \\ \hline
		   $-6.809 \times 10^{-5}$  & $9.442 \times 10^{-1}~~~$
          & $-9.210 \times 10^{-5}$  & $\phantom{-}9.456 \times 10^{-1}$\\
        
         $\phantom{-}6.809 \times 10^{-5}$  & $9.442 \times 10^{-1}~~~$
          & $\phantom{-}9.210 \times 10^{-5}$  & $\phantom{-}9.456 \times 10^{-1}$\\
        
         $-1.159 \times 10^{-4}$  & $9.427 \times 10^{-1}~~~$
          & $-1.136 \times 10^{-4}$  & $\phantom{-}9.326 \times 10^{-1}$\\
        
         $\phantom{-}1.159 \times 10^{-4}$  & $9.427 \times 10^{-1}~~~$
          & $\phantom{-}1.136 \times 10^{-4}$  & $\phantom{-}9.326 \times 10^{-1}$\\
        
         $-2.786 \times 10^{-4}$  & $9.263 \times 10^{-1}~~~$
          & $-1.766 \times 10^{-4}$  & $\phantom{-}9.361 \times 10^{-1}$\\
        
         $\phantom{-}2.786 \times 10^{-4}$  & $9.263 \times 10^{-1}~~~$
          & $\phantom{-}1.766 \times 10^{-4}$  & $\phantom{-}9.361 \times 10^{-1}$\\
        
         $-1.762 \times 10^{-2}$  & $6.264 \times 10^{-3}~~~$
          & $-1.654 \times 10^{-2}$  & $\phantom{-}1.346 \times 10^{-2}$\\ 
        
        $\phantom{-}1.762 \times 10^{-2}$  & $6.264 \times 10^{-3}~~~$
          & $\phantom{-}1.654 \times 10^{-2}$  & $\phantom{-}1.346 \times 10^{-2}$\\
        
        $-1.858 \times 10^{-2}$  & $6.886 \times 10^{-3}~~~$
          & $-2.021 \times 10^{-2}$  & $-2.183 \times 10^{-5}$\\
        \hline
	\end{tabular}
	\caption{Eigenvalues and modified chirality $\mathcal{X}$ using eigenvectors of
    $H(m=0)$ for $Q = 2.995$ background $16^3 \times 48$ MILC lattice.}
    \label{table:chirality_q3} 
\end{table}

\begin{figure}[htb]
\includegraphics[width=0.5\textwidth]{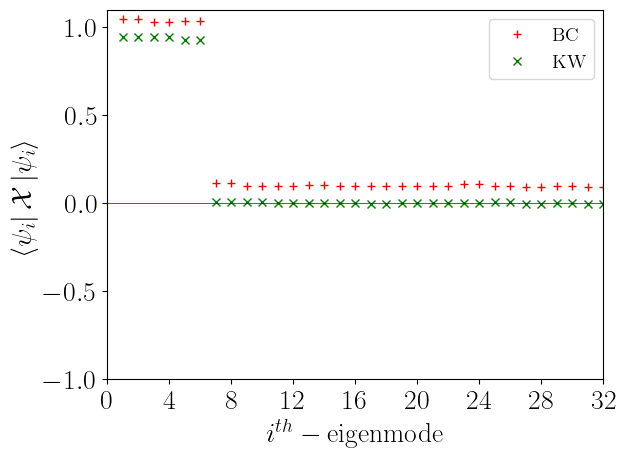}
\vspace{-0.2in}
\caption{The modified chirality $\mathcal{X}$ for $Q=2.995$ background $16^3 \times 48$ MILC lattice.
For clarity, $\langle \mathcal{X}_\text{BC} \rangle$ is shifted by $+0.1$.}
\label{fig:chirality_vs_index_q3}
\end{figure}

\begin{figure}[htb]
\includegraphics[width=0.5\textwidth]{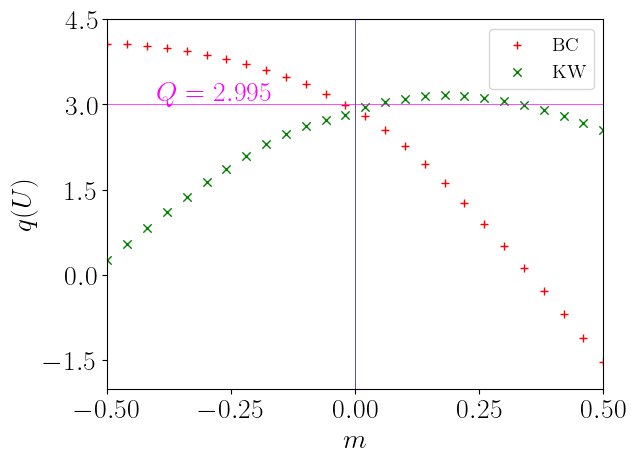}
\vspace{-0.2in}
\caption{Variation of fermionic topological charge with $m$
for $Q=2.995$ background $16^3 \times 48$ MILC lattice.}
\label{fig:fermi_topo_charge_q3}
\end{figure}

\section{Eigenvectors of hermitian matrix \texorpdfstring{$H$}{H} from eigenvectors of Dirac operator $D$ and
their chiralities}\label{appendix_B}
Acting $\gamma_5$ on Eq. (\ref{evDg5D}) from left gives
\begin{eqnarray}
    \implies & \gamma_5 D \ket{\psi_j} & = i \lambda_j \gamma_5 \ket{\psi_j} \label{eq:g5D} \\
    \implies & - D \gamma_5 \ket{\psi_j} & = i \lambda_j \gamma_5 \ket{\psi_j}  \hspace{1cm} \because \{\gamma_5,D\}=0 \nonumber \\
    \implies & D [\gamma_5 \ket{\psi_j}] & = - i \lambda_j [\gamma_5 \ket{\psi_j}]. \label{eq:conj_eigval}
\end{eqnarray}
Hence, if $\ket{\psi_j}$ is an eigenvector of $D$ with eigenvalue $i \lambda_j$, then $\gamma_5 \ket{\psi_j}$ is also an eigenvector of $D$ with eigenvalue $- i \lambda_j$. It is always possible to obtain a space of complete orthonormal eigenvectors $\ket{\psi_k}$ such that $\bra{\psi_k}\ket{\psi_j}=\delta_{kj}$ since the normality condition ($[D,D^\dagger]=0$) holds.
Using the normality of $D$ and its $\gamma_5$-hermiticity, one can choose simultaneous eigenvectors
$\ket{\psi_j}$ of $D$ and $\gamma_5$ in the subspace  $\lambda_j=0$, \textit{i.e.},
$\gamma_5 \ket{\psi_j} =\pm \ket{\psi_j}$ \cite{Kerler:1999pw}.
The proof proceeds as follows. Subtracting Eq. (\ref{eq:conj_eigval}) from Eq. (\ref{eq:g5D}), we obtain
\begin{alignat}{3}
    &  (\gamma_5 D - D \gamma_5) \ket{\psi_j} & = & \quad2i \lambda_j \gamma_5 \ket{\psi_j} \nonumber \\
    \implies \quad & [\gamma_5,D] \ket{\psi_j} = 0 & \iff & \quad \lambda_j=0 \label{eq:commute_g5_D} \\
    \implies \quad & D \ket{\psi_j} = 0 & \& & \quad \gamma_5 \ket{\psi_j} = \pm \ket{\psi_j}. \hspace{1cm} \label{eq:definite_chirality}
\end{alignat}
Now, combining (\ref{eq:definite_chirality}) and $\{\gamma_5, D \}=0$ leads to 
\begin{equation}
\gamma_5 \ket{\psi_j} =
\begin{cases}
    \pm \ket{\psi_j} & \;\; \text{for} \;\; \lambda_j=0 \\
    \phantom{\pm}\ket{\psi_k} & \;\; \text{for} \;\; \lambda_j \neq 0 \;\; \text{with} \;\; \lambda_j=-\lambda_k
\end{cases}. \label{eq:g5_effect}
\end{equation}
Thus, in the space of orthonormal eigenvectors $\ket{\psi_j}$ of $D$, we have
\begin{equation}
    \bra{\psi_j} \gamma_5 \ket{\psi_k} =
    \begin{cases}
        \pm \delta_{jk} & \;\; \text{for} \;\; \lambda_j = \lambda_k = 0 \\
        \phantom{\pm} 1 & \;\; \text{for} \;\; \lambda_j = -\lambda_k \neq 0 \\
        \phantom{\pm} 0 & \;\; \text{for} \;\; \lambda_j \lambda_k = 0 \;\; \text{and} \;\; \lambda_j \neq \lambda_k \\
        \phantom{\pm} 0 & \;\; \text{for} \;\; \lambda_j \lambda_k \neq 0 \;\; \text{and} \;\; \lambda_j \neq - \lambda_k
    \end{cases}. \label{eq:g5_chirality}
\end{equation}
For $H(m)=\gamma_5(D+m)$, we compute its eigenvalues $\mu_j$ and eigenvectors $\ket{\phi_j}$ in terms of the eigenvalues $\pm i\lambda_j$ and the 
corresponding eigenvectors $\ket{\psi_j}$ and $\gamma_5 \ket{\psi_j}$ of $D$. 
\begin{enumerate}
    \item Consider the case with $\lambda_j \neq 0$. Construct $H(m)$ in the subspace spanned by $\ket{\psi_j}$ and $\gamma_5 \ket{\psi_j}$:
    \begin{eqnarray}
        H(m) &=&
        \begin{pmatrix}
            \bra{\psi_j} H \ket{\psi_j} & \bra{\psi_j} H [\gamma_5 \ket{\psi_j}] \\
            [\bra{\psi_j}\gamma_5] H \ket{\psi_j} & [\bra{\psi_j}\gamma_5] H [\gamma_5 \ket{\psi_j}]
        \end{pmatrix} \nonumber \\
        &=&
        \begin{pmatrix}
            0 & -i\lambda_j + m \\
            i\lambda_j + m & 0 \\
        \end{pmatrix}.
    \end{eqnarray}
    Its eigenvalues and eigenvectors are 
    \begin{eqnarray}
    \ket{\phi_j} = \frac{1}{\sqrt{2}} \Big[ \ket{\psi_j} \pm \frac{i \lambda_j+m}{\sqrt{\lambda_j^2 + m^2}} \gamma_5\ket{\psi_j} \Big] \;\;  \text{with} \;\;    \mu_j = \pm \sqrt{\lambda_j^2+m^2}.\label{eq:exact_h_eigenvalue}
	\end{eqnarray}
    Following Eq. (\ref{eq:g5_chirality}), for $m=0$, the chiralities vanish, \textit{i.e.},
    \begin{equation}
        \bra{\phi_j} \gamma_5 \ket{\phi_j}=0.
        \label{eq:non_zero_eigenmode_H_chirality}
    \end{equation}

    \item Consider the case with $\lambda_j=0$:
    \begin{equation}
        H\ket{\psi_j} = \gamma_5 (D+m)\ket{\psi_j}
        = m\gamma_5 \ket{\psi_j}
        = \pm m\ket{\psi_j}. \hspace{1em} 
        \label{eq:D_H_eigvec}
    \end{equation}
    Hence, eigenvectors $\ket{\psi_j}$ of $D$ corresponding to zero eigenvalues $(\lambda_j=0)$, are also eigenvectors of $H(m)$ with eigenvalues $\mu_j=\pm m$ depending upon their chiralities.

    Suppose there are two zero eigenvectors, $\ket{\psi_{j_1}}=\ket{\phi^{+}}$ and $\ket{\psi_{j_2}}=\ket{\phi^{-}}$, with
    positive and negative chirality, respectively. From these, one may construct two orthonormal zero eigenvectors
    of $D$ that are not necessarily eigenvectors of $\gamma_5$:
    \begin{eqnarray}
        \ket{\zeta_1} &=& a_1 \ket{\phi^{+}} + b_1 \ket{\phi^{-}}, \hspace{1em} \text{with} \;\;\; |a_1|^2 + |b_1|^2 = 1, \nonumber \\
        \ket{\zeta_2} &=& a_2 \ket{\phi^{+}} + b_2 \ket{\phi^{-}}, \hspace{1em} \text{with} \;\;\; |a_2|^2 + |b_2|^2 = 1, \nonumber
    \end{eqnarray}
    where $\bra{\zeta_1} \ket{\zeta_2} =0$. The expectation values of $\gamma_5$ are given by:
    \begin{equation}
    \begin{aligned}
        \bra{\zeta_1} \gamma_5 \ket{\zeta_1} = |a_1|^2 - |b_1|^2, \\
        \bra{\zeta_2} \gamma_5 \ket{\zeta_2} = |a_2|^2 - |b_2|^2.
    \end{aligned}
    \label{eq:arb_g5_expec}
    \end{equation}
    Consequently, $\langle \gamma_5 \rangle$ can take any value between $-1$ and $+1$, unless we impose the
    condition that the states be simultaneous eigenvectors of $D$ and $\gamma_5$ within the zero-mode subspace.
\end{enumerate}

\bibliography{index_with_mdf_refs}

\end{document}